\documentstyle [12pt,epsf] {article} 

\begin{document}
\begin{titlepage}
\begin{center}
\vspace{2cm}
\LARGE
Clustering of Galaxies in a Hierarchical Universe:\\  I. Methods and Results at $z=0$   
\\                                                     
\vspace{1cm} 
\large
Guinevere Kauffmann, J\"{o}rg M. Colberg, Antonaldo Diaferio \& Simon D.M. White  \\
\vspace{0.5cm}
\small
{\em Max-Planck Institut f\"{u}r Astrophysik, D-85740 Garching, Germany} \\
\vspace{0.8cm}
\end{center}
\normalsize
\begin {abstract}
We introduce a new technique for following the formation and evolution of galaxies
in cosmological N-body simulations. Dissipationless simulations are used to track the formation and
merging of dark matter halos as a function of redshift. 
Simple prescriptions, taken directly from semi-analytic models of galaxy formation, are adopted for gas cooling, 
star formation, supernova feedback and the merging of galaxies within the halos.
This scheme enables us to                                                                               
explore the clustering properties of galaxies and 
to investigate how  selection  by luminosity, colour or type influences the results.
In this paper, we study the properties of
the galaxy distribution at $z=0$. These include  B and K-band luminosity functions,
two-point correlation functions, pairwise peculiar velocities, cluster mass-to-light ratios, 
$B-V$ colours and star formation rates.                
We focus on two variants of a cold dark matter (CDM) cosmology:
a high-density ($\Omega=1$) model with shape-parameter $\Gamma=0.21$ ($\tau$CDM), and a low-density model  
with $\Omega=0.3$ and $\Lambda=0.7$ ($\Lambda$CDM). Both models are normalized to reproduce the                    
I-band Tully-Fisher 
relation of Giovanelli et al. (1997) near a circular velocity of 220 km s$^{-1}$.                  
Our results depend {\em strongly} both on this normalization and on the adopted prescriptions for star formation
and feedback. Very different assumptions are required to obtain an acceptable model in the two cases.
 For $\tau$CDM, efficient feedback is required to
suppress the growth of galaxies, particularly  in low-mass field  haloes. Without it, 
there are too many galaxies            
and the correlation function
exhibits a strong turnover on scales below 1 Mpc. For $\Lambda$CDM, feedback must be 
weaker, otherwise, too few         
$L_*$ galaxies are produced  and the  correlation function is too steep.
Although neither model is perfect both come close to reproducing most of the data.
Given the uncertainties in modelling some of the critical physical processes, we conclude that it is not
yet possible to
draw firm conclusions about the values of cosmological
parameters from studies of this kind. 
Further observational work on global star formation and feedback effects is required to narrow the range
of possibilities.
\end {abstract}
\vspace {0.8cm}
\end {titlepage}

\section {Introduction}          
A major motivation for carrying out N-body simulations of large-scale structure formation is
to test theories for the origin of structure and to estimate  cosmological
parameters, such as the density $\Omega$, or the cosmological constant $\Lambda$.
Physically accurate identification of the positions, velocities and intrinsic properties of galaxies is
necessary  if such simulations are 
to provide estimates of statistics such as the
spatial and velocity correlations of galaxies, which can be reliably compared with
observations.

Dissipationless simulations of gravitational clustering using tens of millions of particles are
now carried out routinely on parallel supercomputers. Such simulations are able to resolve
the formation and evolution of the  dark matter halos of typical galaxies over cosmologically
significant volumes. However, once these galaxy-sized halos  merge to form larger structures, such
as groups or clusters, they are quickly disrupted and are no longer distinguishable
as separate entities within the more massive systems. This is commonly referred to as the
``overmerging problem''. Several recent papers 
have demonstrated that with sufficient  force and mass resolution  
($\sim 1-3$ kpc and $10^{8}-10^{9} M_{\odot}$) the central cores of many simulated
galaxy halos do ``survive'' in groups and clusters, but 
even with this resolution, substructure is still
erased in the dense central regions
(Tormen, Diaferio \& Syer 1998; Ghigna et al 1998;
Klypin et al. 1998). 

Dissipationless simulations do not address the fact that on scales of a few kiloparsecs, gas-dynamical
processes are known to play a key role in determining the structure of galaxies. 
Numerical simulations that treat the physics of the baryonic component can be used to 
overcome this deficiency. Indeed, it has been demonstrated that when radiative cooling is included,
many individual dense knots of cold gas do survive within a present-day cluster
(e.g. Carlberg, Couchman \& Thomas 1990; Katz, Hernquist \& Weinberg 1992; Navarro \& White 1994;
Evrard, Summers \& Davis 1994; Frenk et al. 1996). 
Such simulations require a heavy investment of CPU time and with present-day computer technology, it is
difficult to follow the formation of galaxies in volumes that are
large enough to study large-scale structure. In addition, the galaxies themselves are very poorly resolved
and processes such as star formation
and supernova feedback 
still have to be put into the simulations ``by hand'' because they are strongly influenced by structure much smaller 
than the resolution limit.

In this paper, we introduce a technique for following the formation and evolution of galaxies
in cosmological N-body simulations. Dissipationless simulations are used to track the formation and
merging of dark matter halos as a function of redshift. The  most bound particle in each halo
is identified as the site where cold gas condenses and forms stars. Later, when two or more
halos merge in the simulation, these particles maintain their identities as
separate galaxies except they can  merge with the new central object  on a timescale set by dynamical friction.
Simple prescriptions are adopted for gas cooling rates, star formation and  supernova feedback.
These are based on simplified models of the physical processes or on simulation results and are
are taken directly from earlier   {\em semi-analytic} studies of galaxy
formation. These used analytic models rather than N-body simulations  to 
specify the merging histories of dark matter halos
for a given set of cosmological initial conditions. Such semi-analytic models were able to 
account for many aspects of the present-day galaxy population, for example the luminosities,
colours and morphologies of galaxies  and the observed correlation of these properties with environment 
(White \& Frenk 1991; Lacey \& Silk 1991; Kauffmann, White
\& Guiderdoni 1993; Cole et al 1994; Baugh, Cole \& Frenk 1996b; Somerville \& Primack 1998).
In addition, semi-analytic models have been used to study the evolution of the galaxy population
to high redshift and to make predictions for faint galaxy counts,
for the redshift distributions of faint galaxies, for the evolution of the morphology-density relation and for the
abundance and properties of galaxies and damped Ly$\alpha$ absorbers 
at high redshifts ( Lacey et al. 1993;
Kauffmann, Guiderdoni \& White 1994;
Heyl et al 1995; Kauffmann 1995a,b; Baugh, Cole \& Frenk 1996a; Kauffmann 1996a,b; Kauffmann \& Charlot 1998;
Baugh et al 1998; Mo, Mao \& White 1998).

The combination of semi-analytic galaxy formation models with cosmological N-body simulations
results in an important advance. It becomes possible to track the positions and velocities of
galaxies as a function of time, in addition to properties such as stellar mass, luminosity,
colour and morphology. One can then compute the standard statistical measures used to
quantify the clustering of galaxies, such as the two-point  correlation function or the
pairwise peculiar velocities, and investigate how the selection of galaxies by luminosity,
colour or  type influences the results. One can extract mock catalogues from the simulations 
in order to study the clustering properties of galaxies in redshift space. One can investigate
whether the galaxy distribution is {\em biased} relative to that of the dark matter and how
this bias affects attempts to measure the density parameter $\Omega$ from galaxy surveys.
Finally, one can also study how the clustering of galaxies
evolves with redshift.

Some, but not all of these issues have been explored in previous papers.
White et al (1987) assumed that a galaxy formed at the centre of a halo during its initial collapse and then
tracked the positions of the central particles as a function
of time. The circular velocities of galaxies  
were given by the the circular velocities of the halos in which they formed. 
Galaxies were also able to merge if they approached within a certain critical radius of each other.
White et al (1987) demonstrated that at the present day, galaxies with high circular velocities cluster
more strongly than galaxies with low circular velocities.
Van Kampen \& Katgert (1997)  used a similar approach to follow the formation of galaxies
within N-body simulations of cluster formation. In their scheme, each group of bound particles
in virial equilibrium was replaced by a single massive  particle with a softening corresponding to the radius
of the group. The studies of White et al and Van Kampen \& Katgert
did not include any scheme for star formation and no attempt was made to compute
the luminosities or colours of the galaxies
in the simulation. 

Modelling of cooling, star formation, feedback and stellar evolution is included
in the cosmological Eulerian grid calculations of Cen \& Ostriker (1992,1993). Even though these
calculations have insufficient resolution to follow the collapse and dynamics of  galaxy halos, and
so cannot reliably identify individual galaxies,
Cen \& Ostriker were able to address some of the issues we discuss in the present paper (e.g. bias and
the dependence of clustering on galaxy age). A number of
their conclusions prefigure our own.

The results of semi-analytic models have
been used  to place galaxies within individual outputs of N-body simulations with higher resolution.   
Analytic methods are used  to compute the luminosity function of galaxies
within  dark matter halos of given  mass $M$ and galaxies are then assigned randomly
to particles in each simulated halo of this mass. 
Kauffmann, Nusser \& Steinmetz (1997) used this method to study present-day bias as
a function of galaxy luminosity, colour and type, while  Governato et al. (1998)
used the same technique to analyze the clustering properties of Lyman break galaxies at redshifts $\sim 3$.
The disadvantage  is that the properties of the galaxies within each simulated halo do not depend on
the merging history of that particular 
halo. Correct results are only obtained if an average is taken over many halos.
Furthermore there is no detailed correspondence between the galaxy populations assigned
to the simulations at different times so that many questions about the evolution of galaxy
clustering cannot be directly addressed.
Finally, the method is subject to all the inaccuracies of the
Press-Schechter theory and its extensions (see section 5 below).
It is therefore advantageous to compute the merging histories of dark halos
directly from the simulations. This has been done by Roukema at al. (1997)
who studied the evolution of both halos and  galaxies  in simulations with scale-free initial
conditions. Their method included  prescriptions for star formation and they were able to demonstrate that  
different assumptions could strongly  affect the predicted number density of low-luminosity galaxies.

This paper is the first in a series studying the properties of galaxies identified
in N-body simulations with cold dark matter (CDM) initial conditions. 
These simulations are much larger than those analyzed in previous work and are
able to resolve the detailed  merging history of the dark matter halo of an  $L_*$  galaxy within a
volume of $\sim 10^7$ Mpc$^{3}$.
We first describe the simulations and
the methods used to construct halo merging trees. We outline the prescriptions adopted to model
cooling, star formation, supernova feedback, galaxy-galaxy merging and the evolution of the
stellar populations within galaxies. We show how the
luminosity functions of galaxies in the simulations compare with  
those derived from the analytic merging trees of Kauffmann \& White (1993). 

We then study the properties of
the galaxy distribution at $z=0$. These include the I-band Tully-Fisher relation,
B and K-band luminosity functions,
the two-point spatial  correlation function $\xi(r)$, two-point velocity correlations, cluster $M/L$ ratios,
the $B-V$ colour distributions of galaxies of different types, and  the predicted $H\alpha$ luminosity
function. We focus on two variants of a cold dark matter (CDM) cosmology:
a high-density ($\Omega=1$) model with shape-parameter $\Gamma=0.21$ ($\tau$CDM), and a low-density model  
with $\Omega=0.3$ and $\Lambda=0.7$ ($\Lambda$CDM). The normalization $\sigma_8$ is chosen to
match the observed abundance of rich clusters
in the Universe at $z=0$. The parameters controlling star formation and supernova feedback are set so that
galaxies with circular velocities of
$220$ km s$^{-1}$ have an I-band magnitude consistent with the 
Tully-Fisher relation of Giovanelli et al (1997).

We show that different ways of treating star formation and feedback in the models have a strong influence
both on the galaxy luminosity function and on  the slope and amplitude of the two-point
correlation function. 
In order for the high-density $\tau$CDM model to come
close to matching the observations, we have to suppress the formation of
galaxies in low-mass halos in the field. This is achieved by assuming that supernova feedback is so
efficient that it can eject a substantial fraction of the baryons 
from the potential wells of low-mass dark matter halos.  Even so, the model still produces an excess of
very bright galaxies in groups and clusters.
The low-density $\Lambda$CDM model produces {\em too few} star-forming field galaxies, even if
feedback is weak and gas never escapes from dark halos. 
We have also investigated the effect of dust extinction on predictions
for the colour distributions and clustering properties of galaxies.
Given the uncertainties in modelling some of the critical physical processes,
we conclude that we cannot reliably  constrain the values of cosmological parameters
using the properties of the galaxy distribution at $z=0$. 

\section{The simulations}

The simulations we use were run as a part of the GIF project, 
a joint effort of astrophysicists from Germany and
Israel. Its primary goal is to study the formation and evolution of
galaxies in a cosmological context using semi-analytical galaxy
formation models embedded in large high-resolution $N$-body
simulations. 
The code used for the GIF simulations is called Hydra. It is a
parallel adaptive particle-particle particle-mesh (AP$^3$M) code (for
details on the code, see\ Couchman, Thomas, \& Pearce 1995; Pearce \&
Couchman 1997). The current version was developed as part of the VIRGO
supercomputing project and was kindly made available by them for the
GIF project. The simulations were started on the CRAY T3D at the
Computer Centre of the Max-Planck Society in Garching (RZG) on 128
processors. Once the clustering strength required a finer base mesh and so an even larger
amount of total memory, they were transferred to the T3D at the
Edinburgh Parallel Computer Centre (EPCC) and finished on 256
processors.

A set of four simulations with $N=256^3$ and with different
cosmological parameters was run (Table 1). All the simulations are ``cluster normalized''.
White, Efstathiou, \& Frenk (1993) introduced
this way of fixing the amplitude by determining $\sigma_8$, the square
root of the variance of the density field smoothed over
$8\,h^{-1}$\,Mpc spheres, such that the observed abundance of high-mass
clusters is matched. More recent determinations of $\sigma_8$ using the observed
cluster X-ray temperature function (Eke, Cole, \& Frenk
1996; Viana \& Liddle 1996) yield similar results. For the low-density
GIF simulations, the result by Eke et al.\ (1996) was taken. For the
$\Omega=1$ simulations, slightly larger values than suggested by Eke
et al.\ (1996) were adopted, according to the earlier result by White
et al.\ (1993).    

The parameters shown in Table 1 were chosen not only to
fulfil cosmological constraints, but also to allow a detailed study of
the clustering properties at very early redshifts. The masses of
individual particles are $1.0\times10^{10}\,h^{-1}\,M_{\odot}$ and
$1.4\times10^{10}\,h^{-1}\,M_{\odot}$ for the high- and low-$\Omega$
models, respectively. The gravitational softening was taken to be
$30\,h^{-1}\,{\rm kpc}$. Gravitational lensing by clusters in this set of four models
has been studied by Bartelmann et al (1998).

In this paper we focus on two of the models: $\tau$CDM and $\Lambda$CDM. 
Both have shape parameter $\Gamma=0.21$.
A value of $\Gamma=0.21$ is usually preferred by analyses of galaxy clustering,
for example  Peacock \& Dodds (1994). This is achieved in the $\tau$CDM model,
despite $\Omega=1$ and $h=0.5$, by assuming that a massive neutrino
(usually taken to be the $\tau$ neutrino) was present during the very
early evolution of the Universe and came to dominate the energy density for a
short period. It then decayed into lighter neutrinos which are still
relativistic, thus
delaying the epoch when matter again started to dominate over radiation.
The neutrino mass and lifetime are chosen such that
$\Gamma=0.21$. For a detailed description of such a model see White,
Gelmini, \& Silk (1995). 

\section {Construction of Halo Merger Trees and Identification of Galaxy Positions}
In order to follow the merging history of dark matter halos in the simulations, we store 
particle positions and velocities at 50 different output times, spaced in equal logarithmic intervals
in redshift from $z=20$ to $z=0$. The construction of merger trees from these simulation outputs
involves the following steps.

A friends-of-friends group-finding program is used to locate virialized halos.
We adopt a linking length which is 0.2 times the mean interparticle separation. Only halos containing
at least 10 particles are included in our halo ``catalogues''. Tests show that
10-particle halos are stable systems. More than 95 \% of  10 particle halos identified at one output
time  are still  
located within groups of 10 particles or more at subsequent times. By ``located", we mean that more than 80\%
of the particles in one  halo are present in the {\em same halo} at the later time.
Halos with masses below about 7 particles do not survive over many output times according to this
criterion.

The lowest luminosity galaxy that we are able to resolve in these simulations thus corresponds to
a galaxy in a halo with 10 particles or  $\sim 10^{11} h^{-1} M_{\odot}$. In these models,
the Milky Way halo has a mass of $\sim 2 \times 10^{12} h^{-1} M_{\odot}$, so the faintest resolved galaxies
are roughly a tenth as bright (i.e. comparable in luminosity to the Large Magellanic Cloud).

We then compute a set of physical quantities for all the halos in the catalogues. These are:
\begin {enumerate}
\item {\em the central particle index.} This is the index of the most-bound particle in the
 halo. This particle has particular significance as it marks the position of the {\em central
 galaxy} of the halo, ie the galaxy onto which gas cools and where it  forms stars.
\item {\em $R_{vir}$}, the virial radius,  defined as the distance from
the central particle within which the overdensity of dark matter is 200 times the background
density.
\item {\em $M_{vir}$,} the virial mass.  This is the mass of dark matter                  
contained within $R_{vir}$.
\item { \em $V_c$,} the circular velocity ($V_c = (G M_{vir}/R_{vir})^{1/2}$).  
\end {enumerate}

We begin with the first simulation output that has at least one halo with
10 particles or more.  The central particles of these halos  mark the locations of the first galaxies
identified in the simulation. We then go to the next output time and loop through all the halos
in the catalogue, searching for {\em progenitor halos} at the previous time.  
A halo at redshift $z_1$ is defined to be a progenitor of a halo at redshift $z_0 < z$, if
\begin {enumerate}
\item more than half its particles are included in the halo at redshift $z_0$; and 
\item its central particle  is also included in that halo.
\end {enumerate}

The most massive progenitor of a halo has a special status --  the properties of its central
galaxy are transferred to the central galaxy of the new halo. We thus
{\em reposition} the central galaxy  at
each output time.
The particle index corresponding to the central galaxy changes, but its associated mass and luminosity
evolve in a smooth fashion. This procedure ensures that cold gas always settles at the centre of a halo. 
The central galaxies of  less massive progenitors become {\em satellites}. The particle index
of a satellite galaxy then remains fixed until  the present day. 
A satellite is said to ``belong'' to a halo if its
particle index is among those linked together by the groupfinder.

This procedure is repeated at every output time until $z=0$. At each timestep, and for all the
halos in the associated catalogue, we store the following information:
\begin {enumerate}
\item The index of the central particle of the largest progenitor of the halo (zero if
 the halo had no progenitors).
\item The indices of all satellite galaxies contained within the halo (particles that were central
galaxies of smaller halos at earlier times, and that are now incorporated within the present halo)
\end {enumerate}
This is sufficient to follow the evolution and merging of all galaxies within the simulation volume 
using the scheme outlined in the next section.

Occasionally a galaxy identified at one time is not included within any halo at a subsequent time.
This may happen for the following reasons:
\begin {enumerate}
\item A halo that was above the 10-particle resolution limit at one time, may fall below this
limit at a later time. In this case, there is a central galaxy that does not belong to any halo
in the next simulation output.
\item Satellite galaxies are occasionally ``ejected'' out of halos, particular during mergers.
In most cases, the satellite will fall back into the halo at a later time.
\end {enumerate}
We keep an index list of these ``lost'' galaxies, checking at each subsequent  
output time  to see whether they have been
re-incorporated into a halo. If the recovered particle was previously a central galaxy, its
properties are transferred to the central galaxy of the new halo if the difference in mass between
the old and new halos is small (less than a factor 2). 
Otherwise, the recovered particle becomes a satellite
galaxy within the new halo. Recovered satellite galaxies simply join the satellite population
of the new halo. Note that lost galaxies comprise only a few percent of the
total galaxy population at all times. Moreover, most of them are faint.

\section {The Physical Processes Governing Galaxy Formation}
Our treatment of the physical processes governing galaxy formation is very similar to that      
described in Kauffmann, White \& Guiderdoni (1993, KWG) and in White \& Frenk (1991). The
reader is referred to these papers for more detailed discussion and for derivations of some
of the equations. Some of our prescriptions, in particular those for feedback and for galaxy-galaxy
merging, have changed since these papers were published. 
We have also incorporated a number of extra features, such as dust
extinction and starbursts during galaxy-galaxy mergers. For completeness, we now present a summary of
our prescriptions for the various physical processes. A detailed discussion of how these are
implemented is given in section 4.8.

\subsection {Gas Cooling}
We adopt the simple  model for cooling first introduced by White \& Frenk (1991). For simplicity,
dark halos are modelled as isothermal spheres truncated at their virial radius $R_{vir}$.
We assume that the hot gas {\em always} has a distribution that exactly parallels that of the
dark matter. The total mass of hot gas in the halos is given by
\begin {equation} M_{hot} = \Omega_b M_{vir} -M_{*} -M_{cool} -M_{eject}, \end {equation}
where $\Omega_b$ is the baryon density of the Universe, $M_{vir}$ is the virial mass of the
halo, $M_{*}$ is the total mass of stars that have formed in  the halo,
$M_{cool}$ is the total cold gas contained in the halo and $M_{eject}$ is
the mass of gas ejected out of the halo by supernova feedback (see section 4.3).

The gas temperature can then be derived from the circular velocity of the halo using the
equation of hydrostatic equilibrium:
\begin {equation} T= 35.9 (V_c/ \rm{km s}^{-1})^2 K \end {equation}
At each radius in the halo, we define a local cooling time through the ratio of the specific
energy content to the cooling rate $\Lambda(t)$ .
\begin {equation} t_{cool}(r) = \frac {3}{2} \frac {\rho_g(r)}{\mu m_p} \frac {kT}{n_e^2(r) \Lambda(t)}
\end {equation}
where $\rho_g(r)$ is the gas density, $n_e(r)$ is the electron density, $m_p$ is
the proton mass and $\mu m_p$ is the molecular weight of the gas.
In this paper, we do not include chemical enrichment. The gas is assumed
to have solar metallicity at all times and we use the solar metallicity cooling curve
in figure 9.9 of  Binney \& Tremaine (1987). (see Kauffmann 1996b and Kauffmann \& Charlot 1998
for a description of models including chemical evolution).

At a given redshift $z$, a cooling radius $r_{cool}$ can be defined
as the radius within the halo where the cooling time is equal to the age of the Universe.
For the case of an Einstein-de Sitter cosmology,
\begin {equation} t_{cool}(r_{cool})= \frac {2}{3} H_0^{-1} (1+z)^{-3/2}. \end {equation}

At high redshifts, for small halos, and for high gas fractions,  
the cooling radius is larger than the virial radius of the halo.
We assume that in the absence of supernovae,  all the hot gas in the halo 
would settle to the centre on a timescale given by the halo
dynamical time ($R_{vir}/V_c$). We thus write the instantaneous cooling rate as
\begin {equation} \dot{M}_{cool} (V_c,z) = \frac {M_{hot} V_c} {R_{vir}}. \end {equation}.
The gas content of the halo is further affected by feedback (see next section) and by 
the infall of new material, which we determine directly from the merging trees. 

At later times, for larger halos, and for low gas fractions, 
the cooling radius lies inside the virial radius and
the rate at which gas cools is calculated using the equation
\begin {equation} \dot{M}_{cool}(V_c,z) = 4 \pi \rho_g(r_{cool}) r_{cool}^{2} \frac {dr_{cool}} {dt}. 
\end {equation}

As has been shown in KWG and in Aragon-Salamanca, Baugh \& Kauffmann(1998), for massive halos 
the cooling rates given by equation 6
lead to the formation of central cluster galaxies that are  too bright and too blue to be
consistent with observation if the cooling gas is assumed to form stars with a standard
initial mass function. It should be noted that cooling flows of hundreds of solar masses
per year are {\em observed} in a number of clusters (see for example Fabian, Nulsen \& Canizares 1991;
Allen \& Fabian 1997), but that
the fate of the cooling gas remains a
mystery. One hypothesis is that it may condense into cold clouds instead of stars (Ferland, Fabian \&
Johnstone 1994).
In our models, we assume the gas cooling in halos with $V_c > 350$ km s$^{-1}$ does not form visible
stars. Note that this is the lowest circular velocity at which we can suppress 
star formation  in cooling flows without destroying our
fit to the Tully-Fisher relation (see section 6.2).

\subsection {Star Formation}

As in KWG, we adopt a simple star formation law of the form
\begin {equation} \dot {M}_{*} = \alpha M_{cold} / t_{dyn}, \end {equation}
where $\alpha$ is a free parameter and
$t_{dyn}$ is the dynamical time of the galaxy. For a central galaxy in a halo, the dynamical time is given by
\begin {equation} t_{dyn}= 0.1 R_{vir}/V_{c}. \end {equation}
For disk galaxies, this is motivated  by noting that, if gas collapses to a centrifugally supported state
within an isothermal halo while conserving angular momentum, the contraction factor is
$\sim 2\lambda$, where $\lambda$ is the spin parameter of the gas, assumed to be the same as
for the halo. N-body simulations find that $\lambda$ scatters around a value of $\sim 0.05$.
For a satellite galaxy, $t_{dyn}$  is held fixed at the value when the galaxy
was last a central galaxy. 

It should be noted that according to the simple spherical collapse model, the virial radius of a
dark matter halo scales with circular velocity and with redshift as $R_{vir} \propto V_c (1+z)^{-3/2}$,
reflecting the fact that halos are smaller and denser at earlier epochs.
This means that $t_{dyn}$ is independent of the circular velocity of the halo,
but will decrease at higher redshift, so star formation rates are higher in halos of the same cold gas content
at high $z$.

The star formation law in equation 7  has received
considerable {\em empirical} support from a recent study of the star formation
rates and gas masses  in 61 nearby spiral galaxies and 36 ``starburst'' systems by Kennicutt (1997),
who finds that such a law can fit the data over several
orders of magnitude in star formation rate and gas density. In normal spirals, about 10  
percent of the available gas is turned into stars per orbital time.

\subsection {Feedback from Supernovae}

The effect of energy ejected by supernova explosions into the interstellar medium of a galaxy
has profound implications for the observed properties of galaxies. 
As shown by Cole et al (1994) and Somerville \& Primack (1998), strong feedback in
low-mass galaxies is required to fit the flat ($\alpha > -1.3$)
faint-end slope of the galaxy luminosity function. Kauffmann \& Charlot (1998) have demonstrated
that substantial feedback in {\em massive} ($\sim L_*$) galaxies is also needed to fit the observed slope
of the colour-magnitude relation of elliptical galaxies.
Unfortunately, both theoretical and observational understanding of how feedback operates in different types
of galaxies is extremely limited at present, so we have no option other than to experiment with a variety
of different prescriptions in order to see what difference they make to our results.

Using basic energy-conservation arguments, it is possible to estimate how much cold gas could be
reheated to the virial temperature of the halo for a given mass of stars formed in the galaxy.
For  the Scalo (1986) initial mass function we use to model the evolution of the stellar
populations in our galaxies (see section 4.6), the number of supernovae expected per solar mass
of stars formed is $\eta_{SN} = 5 \times 10^{-3} M_{\odot}^{-1}$. The kinetic energy of the
ejecta from each supernova, $E_{SN}$, is about $10^{51}$ erg. If a fraction $\epsilon$ of this
energy is used to reheat cold gas to the virial temperature of the halo, the amount of cold gas lost
in time $\Delta t$ can be estimated as
\begin {equation} \Delta M_{reheat} = \epsilon \frac {4}{3} \frac {\dot{M}_* \eta_{SN} E_{SN}} {V_c^2} 
\Delta t. \end {equation}
In the models, $\epsilon$ is treated as a free parameter.

One major uncertainty is whether the gas heated by supernova explosions will leave the halo. In our previous
work, reheated gas was always retained within the halo. Another possibility is that
reheated gas will be ejected out of halo. In the models of Cole et al (1994), reheated
gas was removed  until the halo grew in mass by a factor of two or more, whereupon
it was added once again to the hot gas component. The re-incorporation of expelled gas at a later time    
ensured that the total baryonic mass in halos was conserved and that the baryon
fraction in clusters was close to the global value.

In this paper, we will experiment with both feedback prescriptions. We call the model in which
reheated gas is always trapped within the halo  the ``retention'' model.
The model in which reheated gas is expelled from the halo will be called the ``ejection'' model. 
According to our star formation prescription, stars form efficiently in low mass halos at
high redshifts. In the ejection model, the energy injected into the ISM by the first generation of star formation
is sufficient to expel most of the gas from the halo. The star formation rate then
drops and the galaxy fades until the gas is re-incorporated on the next collapse. Stars in low-mass
galaxies are thus formed in a series of ``bursts'' associated with each factor 2 doubling in
halo mass. This bursting behaviour is much less pronounced for massive galaxies, since
their potential wells are deeper and much less gas will be expelled (equation 9).

The effect of the two feedback prescriptions on the typical star formation history of a $L_*$ galaxy 
(more specifically, the central galaxy of a halo with $V_c=
220$ km s$^{-1}$) 
is shown in figure 1. As can be seen, the ejection prescription results in both higher and
more irregular rates of star formation at high redshift when the galaxies reside within low--$V_c$
halos. At low redshifts, the halo potential wells are deeper and so                       
more effective
at retaining the gas heated during supernova explosions. The star formation rates for
for the two prescriptions thus do not differ very much at the present epoch.

\begin{figure}
\centerline{
\epsfxsize=9cm \epsfbox{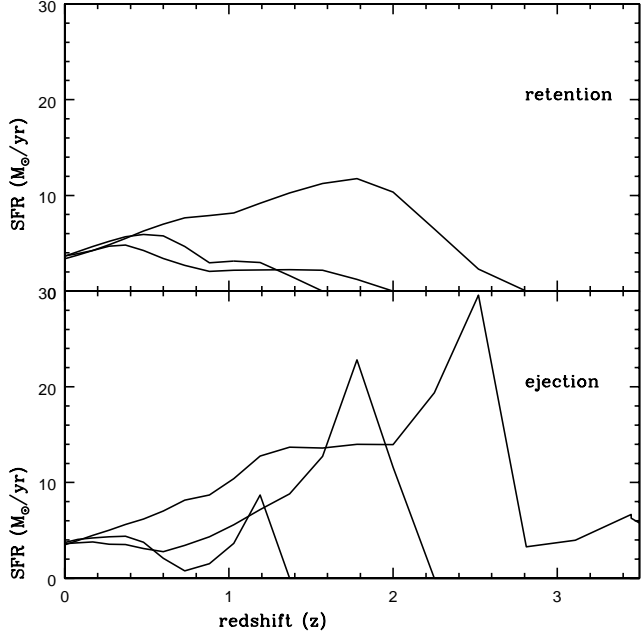}
}
\caption{\label{fig1}
\small
A comparison of the star formation histories for 3 central galaxies in halos of $V_c=220 \quad \rm{km\, s}^{-1}$ for
the retention and the ejection feedback schemes.}
\end {figure}
\normalsize

\subsection {Merging of Galaxies}
N-body plus smoothed particle hydrodynamic (SPH) simulations of the assembly of galaxies in a hierarchical
Universe show that as dark matter halos coalesce, the embedded disk galaxies merge on a timescale
that is consistent with dynamical friction estimates based on their total (gas + surrounding dark matter) mass
(Navarro, Frenk \& White 1995). The dynamical friction timescale, in the form given by
Binney \& Tremaine (1987) is:
\begin {equation} T_{dynf}= \frac{1}{2} \frac {f(\epsilon)} {G C \ln(\Lambda)} \frac {V_c r_c^2}
{M_{sat}}, \end {equation}
where $V_c$ is the circular velocity of a singular isothermal sphere representing the primary halo,
$M_{sat}$ is the mass of the orbiting satellite, $\ln (\Lambda)$ is the Coulomb logarithm which
we approximate as $\ln (M_{vir}/M_{sat})$. The function $f(\epsilon)$ allows for the angular momentum of
the satellite's orbit, expressed in terms of the circularity parameter $\epsilon = J/J_c(E)$, the
ratio of the angular momentum of the satellite to that of a circular orbit with the same energy.
Lacey \& Cole (1993) show that $f(\epsilon) \simeq \epsilon^{0.78}$ for $\epsilon > 0.02$.
Finally, $r_c(E)$ is the radius of the circular orbit with the same energy as the 
satellite orbit.
Navarro, Frenk \& White show that equation 10 provides, on average, a fair estimate of the merger timescale of
a satellite galaxy, provided $M_{sat}$ is taken to be the total gaseous + dark matter mass of the satellite.

In our models, we use equation 10 to compute the timescale for an accreted satellite to
reach the centre of the halo and merge with the central galaxy. The orbital eccentricity parameter $\epsilon$ is drawn randomly
from a uniform distribution from 0 to 1. The radius $r_c$ is set equal to $R_{vir}$, the virial
radius of the primary halo. $M_{sat}$ is taken as the baryonic mass of the satellite plus the mass
of its surrounding halo at the time it was last a central galaxy.

Note that it is assumed that satellite galaxies merge only with the central galaxy of the primary halo
and only after time $T_{dynf}$.
If the primary halo is later accreted by a larger system, new orbital parameters for all the
remaining unmerged satellites
are drawn, the dynamical friction timescales are recomputed and the merger clock is reset to zero.
In practice, substructure in halos will
not be erased immediately and for a while a satellite galaxy may still be able to merge with its old
central object.               

\subsection {Formation of Elliptical Galaxies and Spiral Bulges}
If two galaxies merge and the mass ratio between the satellite and the central object is greater
than 0.3, we add the stars of both objects together and create a bulge component. 
If $M_{sat}/M_{central} < 0.3$, we add the stars and cold gas of the satellite to the disk
component of the central galaxy.
The value of 0.3 is motivated by
a series N-body simulations of merging disk galaxies of unequal mass (Barnes, private
communication). When a  bulge is formed by a merger, all cold gas present in the two 
galaxies is transformed into stars in a ``starburst'' with a timescale
of $10^8$ years. Further cooling of gas in the halo may lead to the formation of a new disk.

The morphological classification of galaxies is made according their B-band disk-to-bulge ratios
(Simien \& de Vaucouleurs 1986). If $M(B)_{bulge}-M(B)_{total} < 1$ mag, then
the galaxy is classified as early-type (elliptical or S0).

It should be noted that although we track the formation of galaxies in halos as small as 
10 particles, we do not accurately predict the morphologies of 
galaxies contained in such halos, simply because their merging histories are not resolved.        
Accurate morphologies are only obtained for central galaxies in halos with $\sim$ 100 particles, 
i.e. for galaxies with luminosities $\sim L_*$.

\subsection {Stellar Population Models}
We use the new stellar population synthesis models of Bruzual \& Charlot (in preparation), which
include updated stellar evolutionary tracks and new spectral libraries. In this paper,
all stars are assumed to have solar metallicity. The star formation
history of any galaxy can be  approximated by a series of delta-function ``bursts'' of
different masses. The stellar population models are used to generate lookup tables of
the luminosity of a burst of fixed mass as a function of age in each photometric band.
The magnitude of the galaxy at $z=0$ is calculated by summing the mass-weighted 
luminosities of each burst.

We have adopted a Scalo (1986) initial mass function with upper and lower mass cutoffs of 100 $M_{\odot}$
and 0.1 $M_{\odot}$. As shown in Kauffmann \& Charlot (1998), the stellar  mass-to-light ratios
of an old (8-10 Gyr) stellar population are then in good agreement with the observed mass-to-light
ratios of elliptical galaxies within an effective radius.

\subsection {Dust Extinction}
In the models there is a strong distinction between central and satellite galaxies. 
Gas cools continuously onto central galaxies and they form stars at a roughly constant rate.
Satellite galaxies lose their supply of new gas and, as a result, have exponentially declining
star formation rates, leading to  redder colours and lower gas fractions than central
galaxies. A central galaxy is thus considerably brighter than a satellite companion
of the same stellar mass at ultraviolet and optical wavelengths.
As we will show in section 6.4, this 
influences the amplitude of the correlation function of galaxies selected according to B-band
magnitude. 

It is a well-known but oft-ignored observational fact that
gas-rich, star-forming galaxies contain dust, which absorbs a substantial fraction of the light
emitted in the short wavelength part of the spectrum and re-radiates it in the far-infrared.
We include a simple, empirically-motivated recipe for dust extinction in our models in
order to investigate how much this effect can influence our estimates of clustering in the models.

Wang \& Heckman (1996) have studied the correlation of the optical depth of dust in galactic disks
with the total luminosity of the galaxy. They studied 150 normal late-type galaxies with 
measured far-ultraviolet (UV, $\lambda \sim 2000 \AA$) fluxes and compiled the corresponding far-infrared
(FIR, $\lambda \sim 40-120 \mu m$) fluxes measured by the IRAS satellite. They then modelled
the absorption and emission of radiation by dust with a simple model of a uniform plane-parallel slab
in which the dust that radiates in the IRAS band is heated exclusively by UV light from nearby
hot stars. They find that their observed UV-to-FIR ratios can be explained by the face-on extinction
optical depth $\tau$ varying with the intrinsic UV luminosity of the galaxy as 
\begin {equation} \tau \propto
\tau_0 (L/L_*)^{\beta}, \end {equation} 
with $\beta \sim 0.5$
The same scaling law was also able to account for the $H\beta/ H\alpha$ ratios
measured for a subset of the galaxies. Re-expressed in the blue-band, Wang \& Heckman derive
a total extinction optical depth of $\tau_{B,*} =0.8 \pm 0.3$ at the fiducial observed blue
luminosity of a Schechter $L_*$ galaxy ($M_*(B)= -19.6 +5 \log h$).

We use the Galactic extinction curve of Cardelli, Clayton \& Mathis (1989) to derive 
$\tau_{\lambda}/ \tau_{B}$.  Although the extinction curves of smaller, less-metal rich galaxies
such as the LMC and SMC are quite different at ultraviolet wavelengths, at optical wavelengths
they are all fairly similar. This leads to face-on extinction for an $L_*$ disk
galaxy of 0.9, 0.8,0.4 and 0.08 mag
in the B,V,I and K bands respectively.

Finally, one must also take into account the inclination of the galaxy to the line-of-sight
when making an extinction correction. For a thin disk where dust and stars are uniformly mixed, the
total extinction in magnitudes is
\begin {equation} A_{\lambda} = -2.5 \log \left ( \frac {1-e^{-\tau_{\lambda} \sec \theta}}
{\tau_{\lambda} \sec \theta} \right ). \end {equation}
In our dust-corrected models, we apply an extinction correction to every galaxy forming stars
at a  rate greater than  0.5 $M_{\odot} \rm {yr}^{-1}$. We use  
$M_*(B)= -19.6+5 \log h - 0.9$ in equation 11,      
and scale $\tau_{\lambda}$ using the {\em uncorrected} B-magnitudes predicted by the model.
We then pick a random inclination for each galaxy and apply the final correction using equation 12.

\subsection {Detailed Implementation of the Prescriptions}     

In the simulation, each galaxy carries a number of ``labels'' corresponding to different physical
properties. These are:
\begin{enumerate}
\item $M_*$, total stellar mass
\item $M_*(bulge)$, stellar mass of the bulge component
\item $M_{cool}$, cold gas mass
\item $L(B,V,R,I,K...)$, total present-day luminosity in a given waveband 
\item $L_{bulge}(B,V,R,I,K...)$, the present-day bulge luminosity  
\end{enumerate}
Satellite galaxies carry two additional labels:
\begin {enumerate}
\item $t_{merg}$, the time until the satellite galaxy should  merge with the central object.
\item $i_{merg}$, the index of the halo in which the satellite resided at the previous timestep.
\end{enumerate}

In the ejection feedback scheme, each {\em halo} carries an array $M_{eject}(M_{prog})$,
which is the amount of gas ejected by galaxies in progenitor halos of mass $M_{prog}$.                

At each output time, we loop over all the halos in the catalogue and follow a series of steps:
\begin{enumerate}

\item We re-incorporate gas that was ejected out of  progenitor halos 
more than a factor of two less massive than the present halo 

\item We calculate the mass of hot gas  in the halo at the start of the timestep. This is given by
\begin {equation} M_{hot} = \Omega_b M_{vir} - \sum_{gals}( M_{cool} +
M_{*}) - \sum_{M_{prog}} M_{eject}(M_{prog}), \end {equation} 
where $\sum_{gals}$ is the sum over all galaxies in the halo.
Note that this  assumes that when
halos merge with each other, all gas that is not already cooled is shock heated
to the virial temperature of the new halo. The sum over $M_{prog}$ extends only over
those progenitors more massive than $M_{vir}/2$.

\item We loop through the satellite galaxies in the halo and update their merging timescales.    
If the satellite was not in the largest progenitor of the halo at the previous timestep, its
merging clock is reset.
If the galaxy is predicted to merge during
the timestep, it is flagged. The time between the beginning of the timestep and the merging
event is noted.

\item We then solve a set of coupled differential equations for the time evolution of the cold gas 
and stars in each of the galaxies in the halo. We adopt a small timestep for this
calculation (there are typically 100 timesteps between each pair of  simulation outputs).
For central galaxies, the change of the cold gas component over time $\Delta t$  is given by:
\begin {equation} M_{cool}(t+ \Delta t)= M_{cool}(t) -( \dot{M}_{*} + \dot{M}_{reheat}- \dot{M}_{cool})\Delta t, \end {equation}
where $\dot{M}_{cool}$, $ \dot{M}_{*}$ and $\dot{M}_{reheat}$ are given by equations 5,6 and 7(or 9) respectively.
For satellite galaxies, the equation is the same   
except there is no source term from gas cooling in the halo. Infall of cold gas onto
satellites is assumed to be {\em disrupted} as soon as they are accreted so that star formation within 
them continues only until their existing cold gas reservoirs are exhausted.

The stars formed during time $\Delta t$  are added to $M_*$. We  look up the present-day luminosity        
of a burst of mass $\dot{M}_{*} \Delta t$ and age $t$ in our population synthesis tables and
update the luminosities of the galaxies. Finally, the hot gas mass of the halo at the end
of the timestep is calculated using
\begin {equation} M_{hot}(t+ \Delta t)= M_{hot} (t) -
\sum_{gals} (\dot{M}_{cool} -\dot{M}_{reheat}) \Delta t. \end {equation}
Note that this is valid for retention feedback. For ejection feedback, reheated gas is added instead to
$M_{eject}(M_{vir})$, where $M_{vir}$ is the virial mass of the halo.

\item
If a satellite is predicted to merge during the timestep $\Delta t$, 
its stars and gas are added to the central galaxy.
If the mass of the satellite is greater than one-third the mass of the central object, we set
the stellar  mass of the bulge component equal to the mass of the
central galaxy plus the mass of the merged satellite. All
the cold gas in the merger remnant is converted into stars at a constant rate over 
$10^8$ years. Stars that form in such a burst  do not reheat or eject cold gas.
In addition, we assume that the the burst is decoupled from the
evolution of rest of the galaxy, i.e. we set $M_{cool}=0$ 
in the above equations immediately  after the merger takes place.
Both assumptions are motivated by N-body simulations, which  show that
gas in merging disk galaxies quickly loses angular momentum and ends up in a dense knot at the
very centre of the remnant, and that feedback has rather little effect (Barnes \& Hernquist 1996). 
\end{enumerate}

\subsection {Normalizing the Models}
For a given set of cosmological initial conditions (including the  baryon density $\Omega_b$), 
there are two ``free parameters'' in our models -- the star formation efficiency $\alpha$ and
the feedback efficiency $\epsilon$. As in  KWG, we tune these two parameters to match the                 
luminosity and cold gas mass of a fiducial reference galaxy, which we take to be the central galaxy
in a halo with circular velocity $V_c= 220$ km s$^{-1}$. In most previous work we have normalized
the models using standard parameters for the Milky Way, but here, following the suggestions of
Somerville \& Primack (1998), we adopt a normalization that is consistent with the velocity-based 
zero-point of the I-band Tully-Fisher relation as determined 
by Giovanelli et al (1997). Matching to I- band rather than B-band data ought to be more
robust, since the I-band is considerably less affected by dust extinction and star formation.
Moreover, a very large amount of data on the I-band Tully-Fisher relation is now available.
Giovanelli et al obtain the following fit, based on 555 galaxies
in 24 clusters:
\begin {equation} M_I -5 \log h = -21.00 \pm 0.02 -7.68 \pm 0.14 (\log W -2.5) \end {equation}

To convert between the measured $HI$ line-widths $W$ and the model circular velocities we assume
$W= 2V_c$. In practice, there is substantial uncertainty in the relation between the circular
velocity of the halo and that of the disk (e.g. Mo, Mao \& White 1998), 
since the transformation between the two quantities
depends on the detailed density profile of the halo and whether or not gas loses angular momentum
before settling onto the disk. We set the parameters $\alpha$ and $\epsilon$ so that the central
galaxy in a halo with $V_c=220$ km s$^{-1}$ has $M_I - 5 \log h \sim -22.1$ and a cold gas mass
$\sim 10^{10} M_{\odot}$. It should be noted that this normalization is considerably
brighter than that adopted by KWG, who took the observed B-band
magnitude of the Milky Way to be $B \sim -20.5$. For a Hubble constant of 50 km s$^{-1}$ Mpc$^{-1}$
and assuming  $B-I \sim 1.8$ for a spiral galaxy (De Jong 1996), the new normalization means that 
the central galaxy in a halo with $V_c= 220$ km s$^{-1}$ has a B-band magnitude of -21.8 -- more
than a factor 3 more luminous than the observed value. Consistency of this Tully-Fisher normalization 
with the estimated magnitudes of our own Galaxy and of M31 is only obtained for substantially higher values
of the Hubble constant, which do not give acceptable ages for a Universe with $\Omega=1$.

The fact that there are only two free parameters in the model should not be interpreted as an indication
that the physical processes governing galaxy formation are well-specified.
On the contrary, we will show that the way in which we choose to implement feedback and whether
or not dust is included in the model can have a very large effect  both on
the luminosity function and on the amplitude of the correlation function on scales below a few Mpc.
We will attempt to clarify how different parametrizations of these processes affect our
results and which processes appear to be most critical if one is to succeed in matching     
the observational data.

\section {Comparison of galaxy properties for semi-analytic and simulation merger trees}
If halo merging trees derived using methods based on the extended Press-Schechter
formalism (Cole 1991; Kauffmann \& White 1993; Rodrigues \& Thomas 1996 Somerville \& Kolatt 1998) 
were statistically equivalent to the trees found in the N-body simulations, 
the same galaxy luminosity functions would be obtained                    
using the two approaches, provided the same recipes for cooling, star formation
and feedback were employed, and the same halo mass resolution limit was adopted.  
In practice, the halo mass function derived by the Press-Schechter argument
does not fit very accurately the mass function in N-body simulations. For halo masses 
between $10^{11}$ and $10^{14}$ $M_{\odot}$,  the Press-Schechter theory predicts roughly twice as many     
halos as are  actually found in the GIF simulations for both
the $\tau$CDM and the $\Lambda$CDM cosmologies. Most of the missing mass in the simulations is in the
form of ``unresolved'' material --  single particles or groups with less than 10 members. 
This discrepancy between the Press-Schechter and simulation mass functions  propagates 
to the halo progenitor distributions and to the Monte-Carlo merging trees 
(Somerville, Lemson \& Kolatt, in preparation).

Here we compare the the predictions of the two approaches for the 
luminosity functions of galaxies within halos of given mass
and for the ``field'' luminosity function.
In figure 2, we plot the number of galaxies brighter than a given V-magnitude
in halos with circular velocity $V_c$. Solid squares show mean results derived using analytic trees
constructed as described in Kauffmann \& White (1993). Solid circles show results derived from
the simulation. The error bars on the points show the rms scatter in the number of
galaxies in halos of given circular velocity. 
We perform the comparison using the $\tau$CDM model and we use the {\em same}  
prescriptions (from KWG) for cooling, star formation, feedback, merging and stellar population synthesis
in both cases. We truncate the analytic trees at a halo mass of $2 \times 10^{11} M_{\odot}$ so
that the resolution is the same as in the simulations. The four panels in figure 2 show the number
of galaxies per halo  brighter than $M_V= -18, -19, -20$ and $-21$.
As can be seen, the mean number of lower-luminosity galaxies per halo  ($M_V > -20$)  
agrees remarkably well. The analytic approach appears to overpredict the number of bright galaxies
in high $V_c$ halos by a factor of $\sim 2$, but since there are not many such halos in the 
simulation and the number of
bright galaxies within them is small, the discrepancy may not be statistically significant.

\begin{figure}
\centerline{
\epsfxsize=12cm \epsfbox{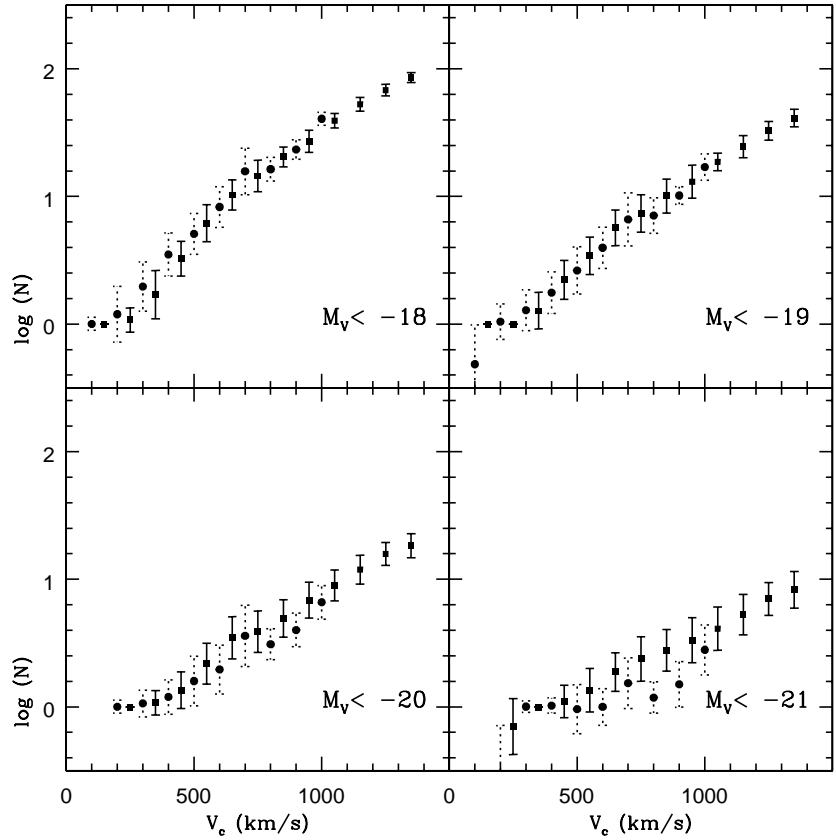}
}
\caption{\label{fig2}
\small
The number of galaxies brighter than a given V-magnitude
in halos with circular velocity $V_c$. Solid circles show results derived using the analytic trees
constructed as described in Kauffmann \& White (1993). Solid squares show results derived from
the simulation.  Error bars show the rms deviation between the number of galaxies in different halos.}
\end {figure}
\normalsize

Figure 3 compares the field luminosity functions derived using the semi-analytic trees and
the simulations. The top panel shows the difference between the abundance of halos predicted 
by Press-Schechter theory,  and that found in the simulation. The Press-Schechter overprediction of halo
abundance at intermediate masses in quite evident. The middle panel 
compares the magnitude of the {\em central} galaxy as a function of halo circular velocity. There
is good agreement between the two approaches.
Note that the dip in magnitude at $V_c= 500$ km s$^{-1}$ arises because
we assume stars do not form in the cooling flows of halos  with circular velocities greater than this value. 
The bottom panel compares the V-band ``field'' luminosity functions. Except perhaps near the ``knee'', the
agreement between the two approaches is excellent.

The comparison in this section demonstrates that although  merging trees derived using the extended
Press-Schechter theory differ in detail from those found in  the simulations, 
the two methods give very similar results both for the mean number and luminosity of
galaxies within halos, and for the scatter in these quantities.  

\begin{figure}
\centerline{
\epsfxsize=12cm \epsfbox{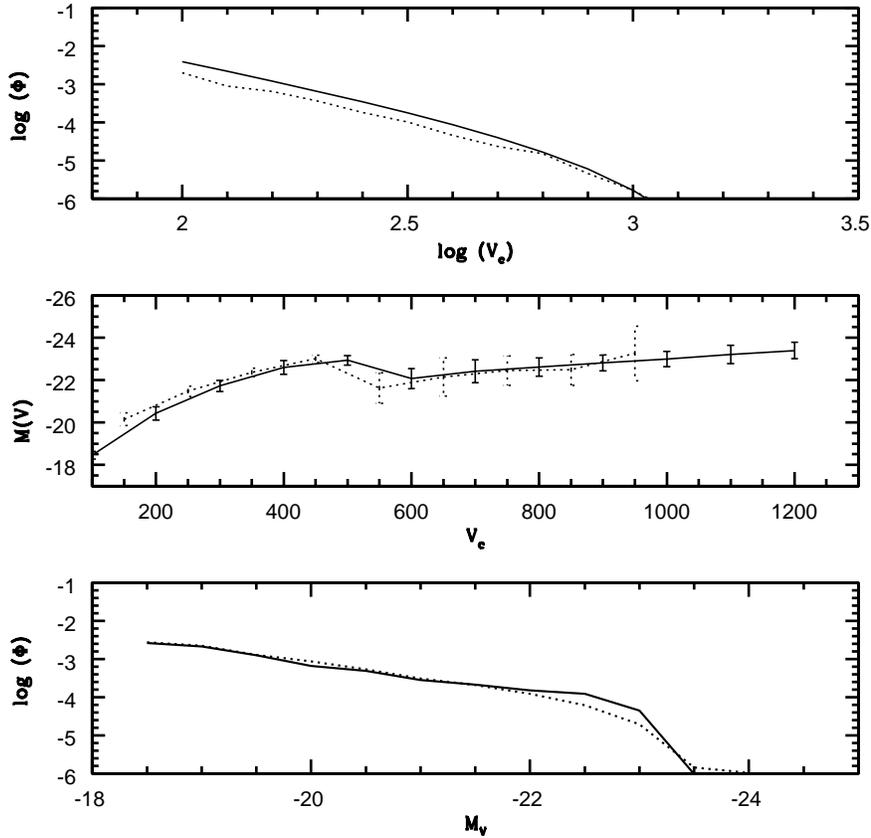}
}
\caption{\label{fig3}
\small
{\em Top panel:} 
A comparison of  the abundance of halos as a function of circular velocity  predicted by the
Press-Schechter theory (solid) and the abundance derived from the 
simulation (dotted). 
{\em Middle panel:}
Magnitude of the {\em central galaxy} as a function of halo circular velocity. Error bars show
the rms scatter between halos. The Press-Schechter theory halos are solid and the simulation is dotted.
{\em Bottom panel:}
The V-band field luminosity functions derived using  the Press-Schechter approach (solid) and
the simulations (dotted).}
\end {figure}
\normalsize

\section { Results of the Models}
\subsection {Slices from the simulation}
In figure 4, we show slices of thickness 8 $h^{-1}$ Mpc from the $\tau$CDM simulation.
The distribution of dark matter is shown in the top left panel. In the top right panel, 
all the galaxies in the slice with magnitudes brighter than $M(B)=-19.0+ 5 \log h$ are plotted
as solid white circles. 
Galaxies with magnitudes brighter than $M(B) = -17.5 +5 \log h$
are shown in the bottom panels. We divide these galaxies into two equal subsamples
at the median star formation rate per unit stellar mass. 
The bottom left panel shows the distribution of galaxies with low star formation rates.         
The bottom right panel shows galaxies with high star formation rates.                           

Galaxies of all luminosities trace the filaments and knots visible 
in the dark matter distribution. Galaxies with  high star formation rates are less clustered than galaxies
with low star formation rates. Star-forming galaxies tend to avoid 
dense clusters and groups and occur more frequently in voids.

Figure 5 shows a similar set of slices from the $\Lambda$CDM simulation. 
The area and thickness of these slices 
have been chosen to have the same dimensions in {\em redshift space} as
those in figure 4 ($85 h^{-1}$ Mpc $ \times 85 h^{-1}$ Mpc $\times 8 h^{-1}$ Mpc). The same
systematic effects are visible in the two figures and the most obvious difference between them is
the difference in galaxy abundance. We return to this in section 6.3.

\begin{figure}
\centerline{
\epsfxsize=17cm 
}
\caption{\label{fig4}
Slices of thickness 8 $h^{-1}$ Mpc from the $\tau$CDM simulation.
The distribution of dark matter is shown in the top left panel. In the top right panel, 
all galaxies with  $M(B)< -19.0+ 5 \log h$ are plotted
as solid white circles. 
Galaxies with  $M(B) < -17.5 +5 \log h$
are shown in the bottom panels. We divide these galaxies into two equal subsamples
at the median star formation rate per unit stellar mass. 
The bottom left panel shows the distribution of galaxies with low star formation rates.         
The bottom right panel shows galaxies with high star formation rates.}                           
\end {figure}
\normalsize

\begin{figure}
\centerline{
\epsfxsize=17cm 
}
\caption{\label{fig5}
Slices of thickness 8 $h^{-1}$ Mpc and length 85 $h^{-1}$ Mpc  from the $\Lambda$CDM simulation.
The panels are as described in figure 4.}
\end {figure}
\normalsize

\subsection {The  Tully-Fisher Relation}
As described in section 4.9, all models are normalized so that the average I-band magnitude of the central
galaxy in a halo of circular velocity $V_c=220$ km s$^{-1}$ is $-22.1+ 5\log h$, in accordance with
the zero point of the  I-band Tully-Fisher relation derived by Giovanelli et al. (1997). 
Figure 6 shows the Tully-Fisher relation for
spiral galaxies in  the simulations. We have selected central galaxies with
$1.5 < M(B)_{bulge}-M(B)_{tot} < 2.2$ (appropriate for Sb/Sc type galaxies).
The solid line is the fit to the data (equation 16).
The upper two panels show the $\tau$CDM model for the two different feedback prescriptions discussed
in section 4.3.
Although the slopes are similar,
the ejection prescription results in considerably more scatter, particularly
at low velocity widths, where there is a marked tail of low-luminosity galaxies.                 
The lower panel shows the $\Lambda$CDM model with retention feedback.                       
We do not show a $\Lambda$CDM model with ejection feedback as this model would result in 
too low a total luminosity density, as explained in the next section. 
It is interesting that the scatter obtained for both the $\Lambda$CDM and $\tau$CDM relations, even for  
retention feedback, is not much smaller than the observed scatter.                              
The model scatter arises because of differences in  star formation
history and stellar mass between  central galaxies in different halos of the same circular velocity.
As pointed out by Mo, Mao \& White (1998), scatter in the spin parameter of
dark matter halos leads to a spread  in  contraction factors for disks forming in halos of
a given  circular velocity. Halos with high values of $\lambda$ produce disks with
rotation curves that are slowly rising, whereas halos with low values of $\lambda$ produce disks with
rotation curves that rise steeply and then decline.
This leads to intrinsic scatter in the predicted Tully-Fisher
relation even if disks in halos of given circular velocity  are assumed to have fixed mass and 
mass-to-light ratio. This must be added to the scatter already visible in figure 6.
More complex modelling is necessary to investigate these issues in detail. 

\begin{figure}
\centerline{
\epsfxsize=12cm \epsfbox{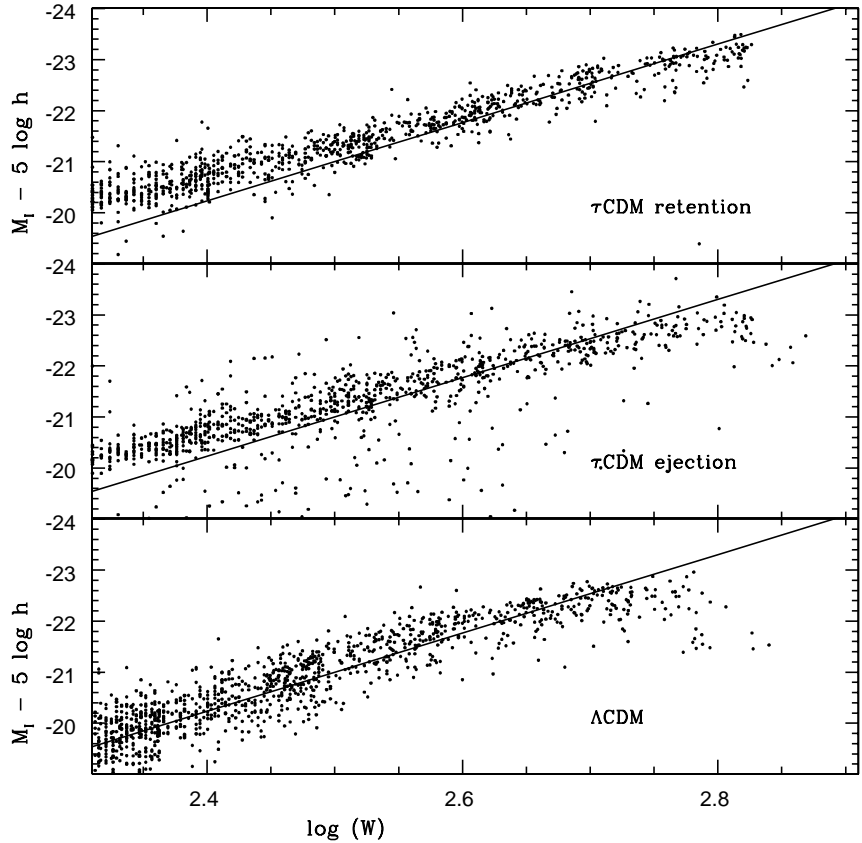}
}
\caption{\label{fig6}
\small
The I-band Tully-Fisher relation for spiral galaxies in the simulations.
The solid lines show the relation of Giovanelli et al (1997).}
\end {figure}
\normalsize

\subsection {The  Luminosity Function}
In figure 7, we show B-band luminosity functions for the $\tau$CDM simulation. A sequence of different models
is shown in order to illustrate the effect of including different prescriptions for feedback and
dust extinction. The results are compared with the B-band luminosity functions derived from
several  recent redshift surveys: the APM-Stromlo survey (Loveday et al 1992),
the Las Campanas Redshift Survey (LCRS) (Lin et al 1997) and  the
ESO-Slice survey (Zucca et al 1997). Note that we only plot the luminosity functions down to magnitudes where 
our simulations are ``complete'' (i.e. the luminosity of the central galaxy in a halo with 10 particles).

In the top left panel, we plot the  luminosity function for a model with retention 
feedback and no dust extinction. In this case, the model luminosity function exceeds the observed one by more than 
a factor of 10 at all luminosities. The inclusion of dust brings the model        
more in line with the observations at luminosities around $L_*$, but there are still far too many galaxies
at both fainter and brighter magnitudes. The model with ejection feedback agrees much better with
the observations at $L_*$ and below. This is because ejection feedback suppresses the formation of bright
galaxies in low mass halos. The first generation of stars formed when the halo collapses  generates
enough energy to empty the halo of gas and halt further star formation. This process is only efficient
in low-circular velocity systems, so the excess of galaxies brighter than $L_*$ still remains a
problem. Including dust in the ejection  model (lower right panel) helps somewhat, but the predicted abundance 
of bright galaxies
is still more than a factor 10 above the Schechter function representation of the observations.

\begin{figure}
\centerline{
\epsfxsize=12cm \epsfbox{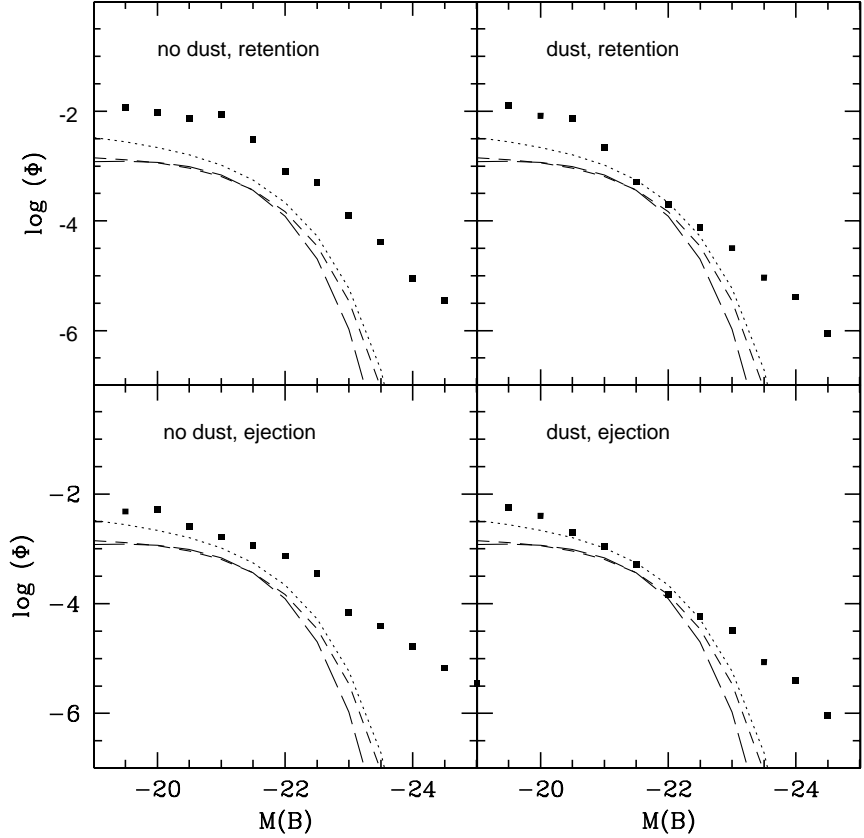}
}
\caption{\label{fig7}
\small
The B-band luminosity function for galaxies in the $\tau$CDM simulation. The four panels show models
with different assumptions about feedback and dust extinction. The simulation results are shown as solid
squares. The lines are Schechter fits to B-band  luminosity functions from  recent redshift surveys:
1) ESO-Slice (dotted), 2) APM-Stromlo (short-dashed), 3) LCRS (long-dashed).}
\end {figure}
\normalsize

In figure 8, we compare the model K-band luminosity functions with   
the Schechter fits derived by Gardner et al (1997) and Szokoly et al (1998) for two
independent surveys. Unlike B-band magnitudes,  the K-band
luminosities of galaxies depend on their total stellar masses, rather than their present-day star formation rates.
Moreover, K-magnitudes are only weakly affected by
dust extinction. The fact that the model K-band luminosity functions also
disagree with the data at the bright end  shows that there are too many {\em massive} galaxies in the simulations.

\begin{figure}
\centerline{
\epsfxsize=10cm \epsfbox{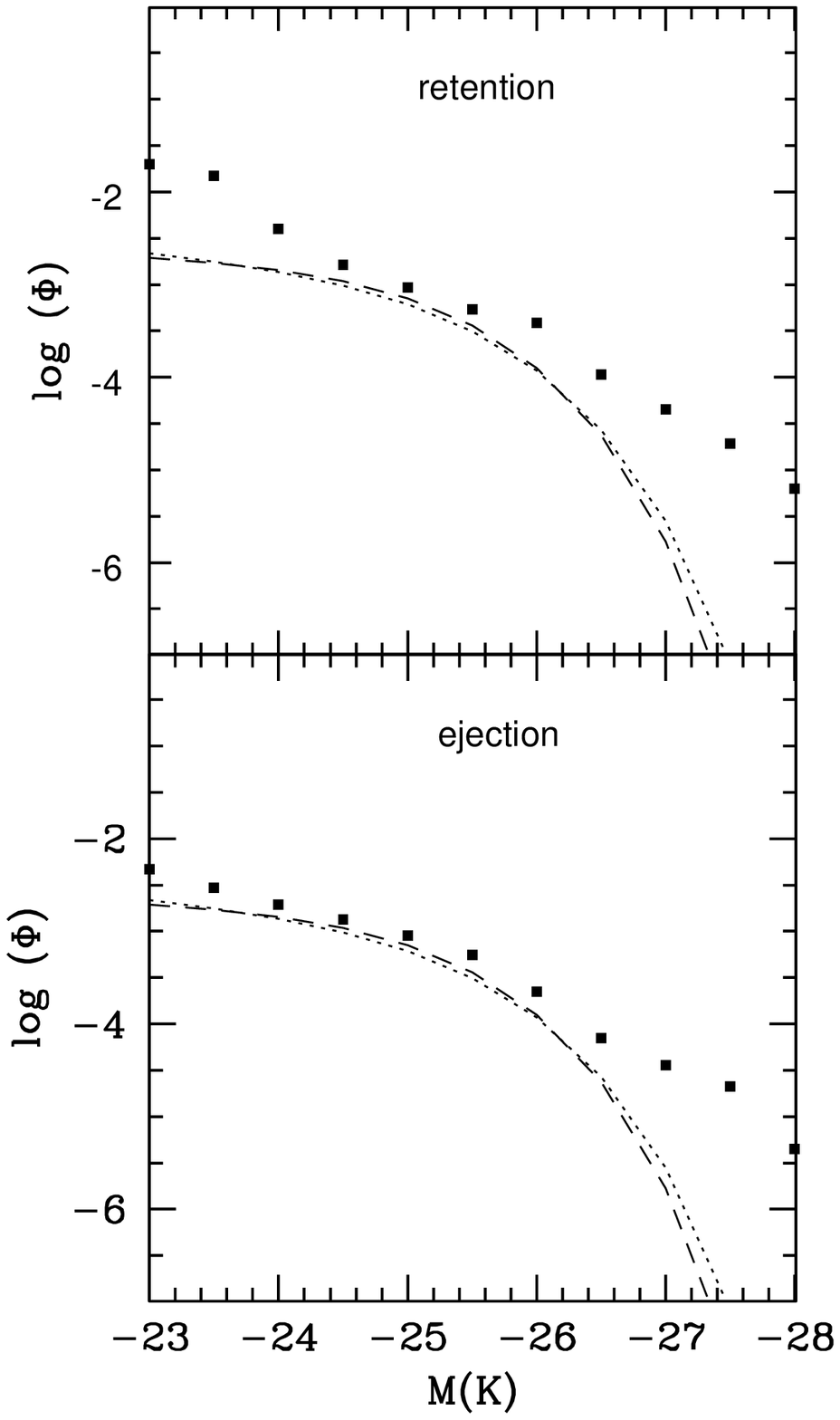}
}
\caption{\label{fig8}
\small
The K-band luminosity function for galaxies in the $\tau$CDM simulation. The two panels show models
with different assumptions about feedback. Dust extinction is negligible in the K-band.
The simulation results are shown as solid
squares. The lines are Schechter fits to K-band luminosity functions derived by                         
Gardner et al (dotted) and Szokoly et al (dashed)}                                     
\end {figure}
\normalsize

B-band and K-band luminosity functions for the $\Lambda$CDM model are shown in figure 9. Retention feedback
is assumed. Even so, this model
produces a factor $\sim 2-3$ too few galaxies at magnitudes around $L_*$.
There is again an excess of very bright galaxies, but the problem is less severe than in $\tau$CDM.
The same trends are apparent in the K-band. 

\begin{figure}
\centerline{
\epsfxsize=12cm \epsfbox{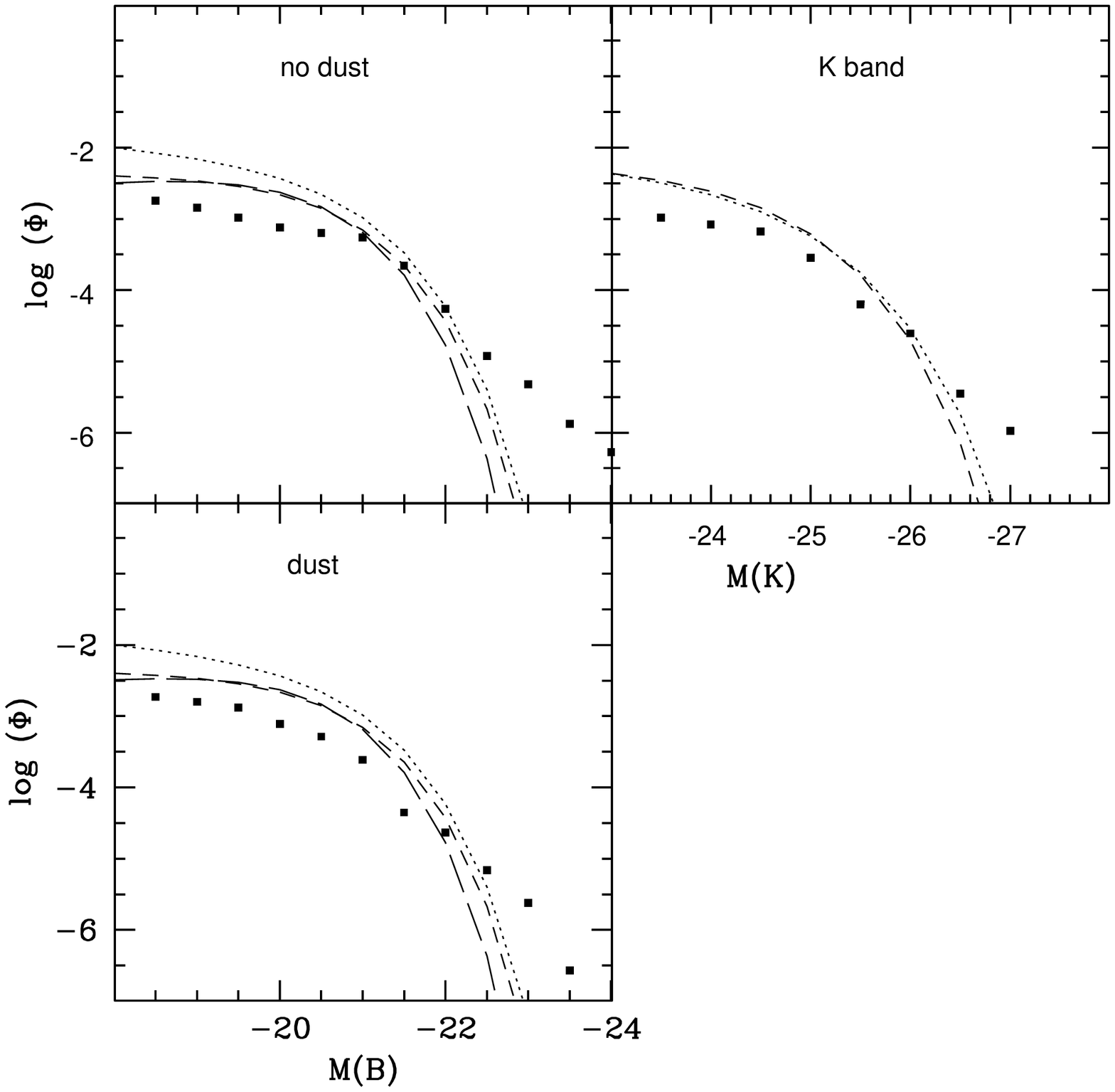}
}
\caption{\label{fig9}
\small
The B and K-band luminosity functions for galaxies in the $\Lambda$CDM simulation. 
The lines are as described in the previous two figures.}
\end {figure}
\normalsize

In summary, the luminosity functions for both models have the wrong shape, with or without
dust extinction. Instead of declining exponentially
at bright magnitudes, they exhibit a gentler turn down. In previous work, KWG and Cole et al (1994) 
obtained much better fits to the bright end of the luminosity function (see also figure 3). 
As explained in section 4.9,
this is because they adopted a much fainter normalization. 
If one is to obtain an exponential cutoff at high luminosities, bright galaxies must reside primarily in
very massive halos on the exponentially declining part of the mass function.
The shape problem is considerably
worse for the $\tau$CDM model than for the $\Lambda$CDM model. Although both models are 
normalized to reproduce the observed abundance of rich clusters, the $\tau$CDM model
has a higher abundance of intermediate mass halos than the $\Lambda$CDM model. 
It is the luminous  galaxies in these halos that create the excess at magnitudes brighter than $L_*$. 
The $\tau$CDM model produces too many  galaxies below $L_*$ unless gas is efficiently
ejected out of low mass halos. The $\Lambda$CDM model, on the other hand, produces too {\em few}
$L_*$ galaxies even if feedback is inefficient and  if gas never escapes these halos.                        

\subsection {The Two-point Correlation Function}
In figure 10, we show the galaxy two-point correlation function 
$\xi(r)$ for the $\tau$CDM simulation. We select galaxies with B-band luminosities $M(B)< -20.5$. 
A sequence of models is again shown  to illustrate different prescriptions for feedback and dust
extinction. The results are compared with the real space correlation function measured from the
APM galaxy survey (Baugh 1996). 

For the model with retention feedback, $\xi(r)$ follows a power law of
slope $\gamma \sim -1.7$
down to scales $\sim 1$ Mpc, below which it flattens and then turns over.
This is in marked contrast to the observed $\xi(r)$, which  follows 
the same power law down to scales below 100 kpc. The inclusion of dust extinction 
lowers the luminosity of bright star-forming field galaxies and raises the amplitude
of the correlation function on small scales, but $\xi(r)$ still flattens and
is a factor 6 below the observed value at 
$r \sim 100$ kpc. As described in the previous section, ejection feedback        
suppresses bright star-forming galaxies in low circular velocity halos
and brings the faint end of the luminosity function into better agreement with the observations.
As seen in the bottom left panel of figure 10, ejection feedback also prevents $\xi(r)$ from flattening below
1 Mpc. The results shown in the bottom right panel of figure 10, for
the model with ejection feedback and dust extinction, agree rather well with the observations.       

\begin{figure}
\centerline{
\epsfxsize=12cm \epsfbox{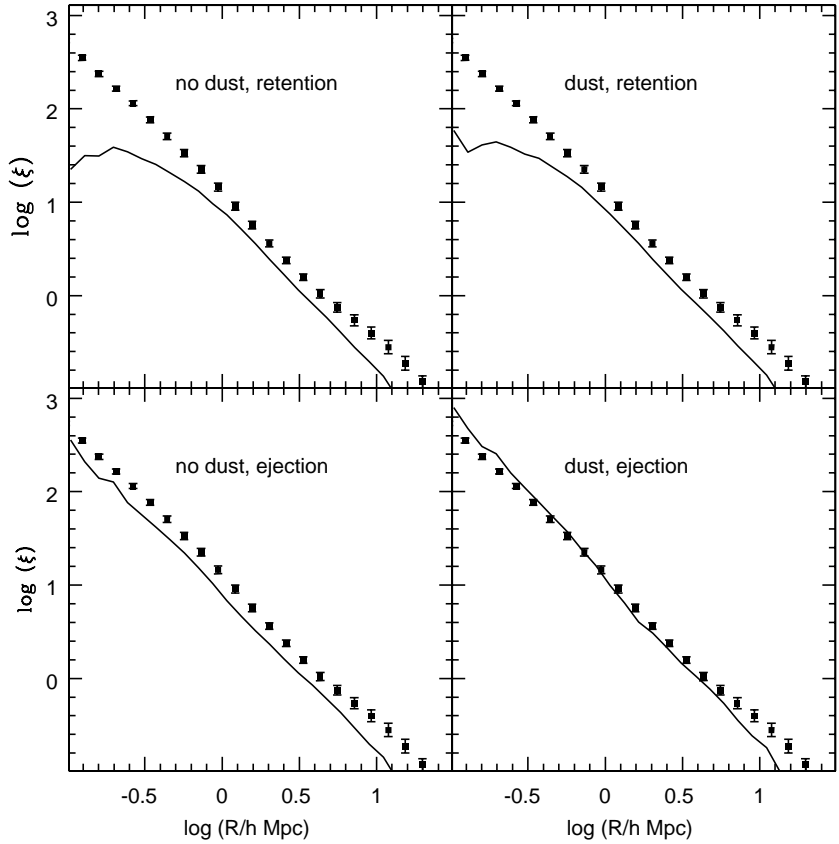}
}
\caption{\label{fig10}
\small
The two-point correlation function $\xi(r)$ for  galaxies in the $\tau$CDM simulation. 
Solid lines show the simulation results for different prescriptions for feedback and
dust extinction.
The points with error bars are taken from Baugh (1995).}
\end {figure}
\normalsize

Figure 11 shows the galaxy two-point correlation function for the $\Lambda$CDM simulation.
Retention feedback is assumed. 
$\xi(r)$ does not turn over on small scales  and the model with dust is actually
{\em too steep} compared to observations.

\begin{figure}
\centerline{
\epsfxsize=9.3cm \epsfbox{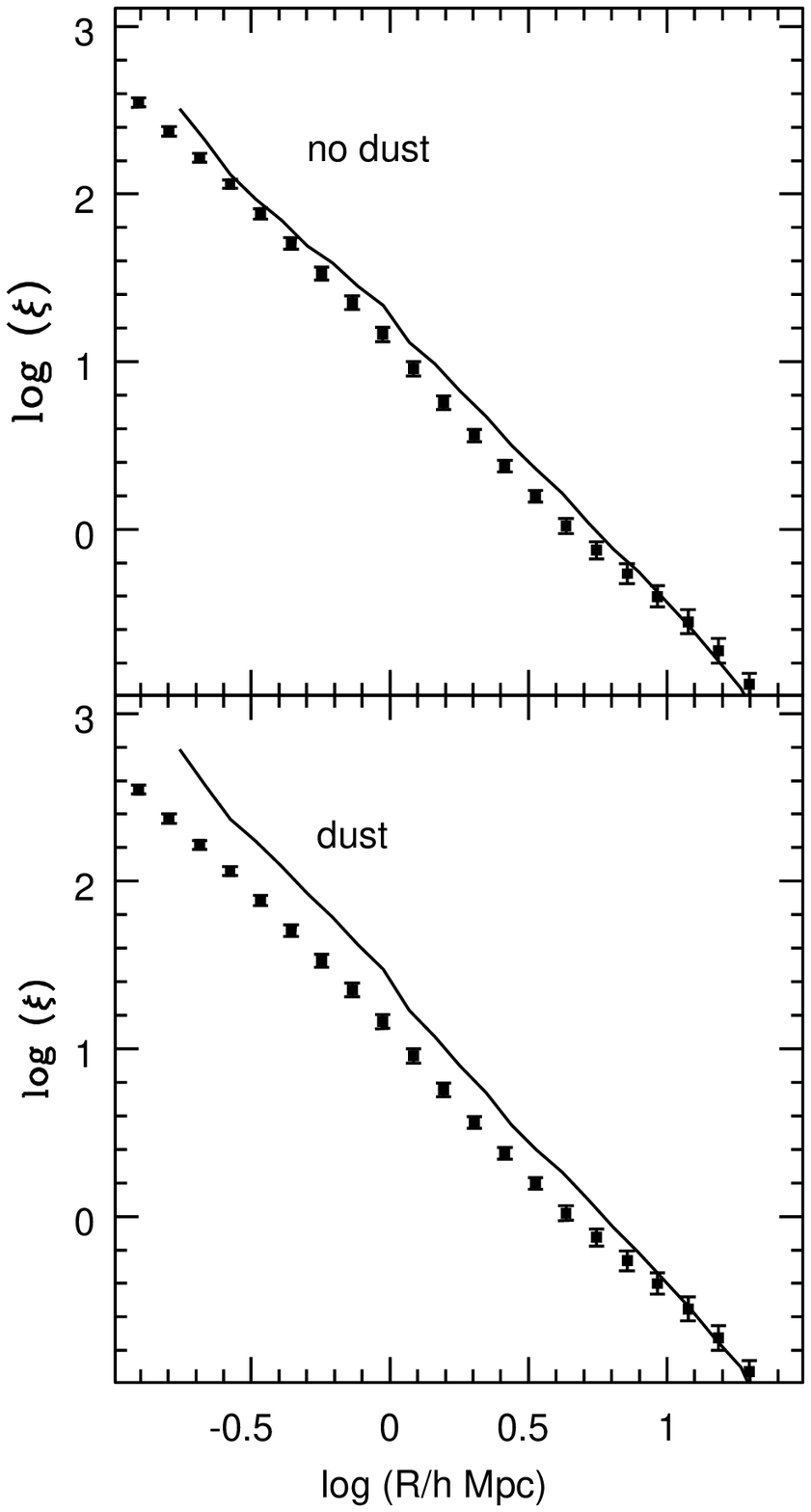}
}
\caption{\label{fig11}
\small
The two-point correlation function $\xi(r)$ for  galaxies in the $\Lambda$CDM simulation (solid line). 
Points are observational data from the APM survey.}
\end {figure}
\normalsize

What determines the slope and amplitude of the galaxy correlation functions in our models?                    
In a recent paper, Mo \& White (1996) developed a simple analytic model for the gravitational clustering
of dark matter halos. The positions and formation times of halos were determined from the
statistics of the initial linear density field and modifications caused by gravitationally-induced motions
were treated using a spherical collapse approximation. Mo \& White showed that on large scales,
the halo autocorrelation function is simply  proportional to the dark matter correlation function, with
the constant of proportionality dependent on the masses of the selected halos. Because halos are
spatially exclusive and have a finite radius, the halo-halo correlation function drops below
the mass correlation function on scales comparable to the diameter of typical halos in the sample.
From these results, it follows that if there is  only one bright galaxy per halo, the galaxy correlation   
function will simply  follow the  form of the halo correlation function, with
a flattening and turnover on scales less than $\sim 1$ Mpc. In order for the galaxy correlation to
continue as a power law to small scales, a substantial fraction of bright galaxies must exist as groups
{\em within the same halo}.

This is illustrated  quantitatively in figure 12, where we plot the fraction
of galaxies in the simulation  with $M(B) < -20.5$ that occur in halos of a given circular velocity. The
$\tau$CDM model with retention feedback has  a much higher fraction of galaxies  in low $V_c$ halos
than the $\tau$CDM model with ejection feedback. Low $V_c$ halos typically contain only one
bright galaxy (see figure 2), whereas high $V_c$ halos contain many bright galaxies that contribute
to clustering amplitude on sub-megaparsec scales. Figure 12  shows that there is a relatively small
fraction of bright  galaxies in low $V_c$ halos in the $\Lambda$CDM model, even with retention feedback.
Recall that both $\tau$CDM and $\Lambda$CDM are normalized to fit the observed
number density of rich clusters at the present day. Because the mass density in the $\Lambda$CDM
model is lower, the number density of low mass halos is also smaller. $\Lambda$CDM thus ``naturally''
has more galaxies in clusters  relative to the field and as a result, $\xi(r)$ is steep on small scales.      

\begin{figure}
\centerline{
\epsfxsize=11cm \epsfbox{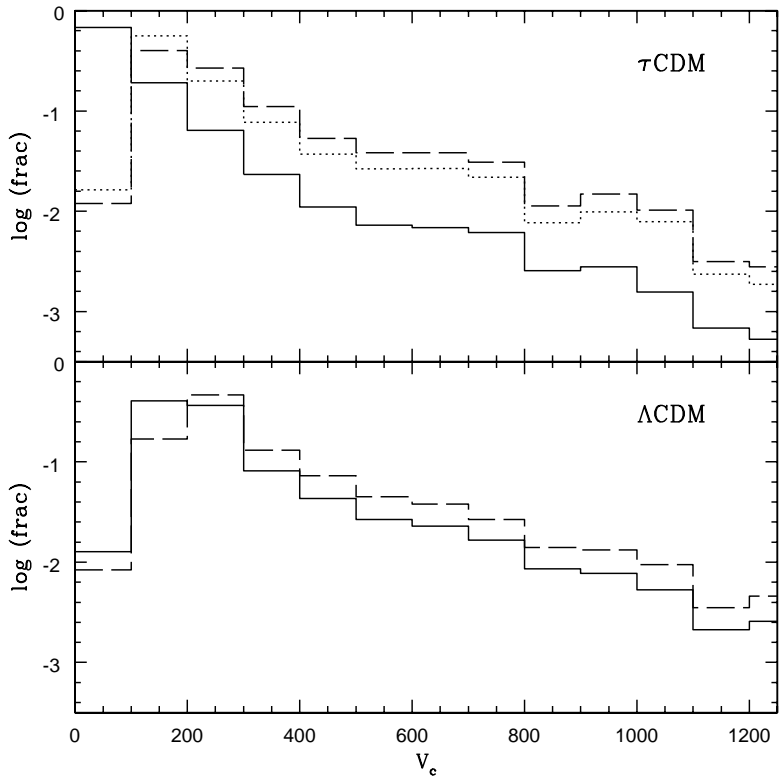}
}
\caption{\label{fig12}
\small
The fraction of galaxies in the simulation  with $M(B) < -20.5$ that occur in halos of a given circular velocity. 
{\em Upper panel:} $\tau$CDM, retention feedback (solid), ejection feedback (dotted),
ejection feedback + dust (dashed).
{\em Lower panel:} $\Lambda$CDM, retention feedback (solid), retention feedback + dust (dashed).}
\end {figure}
\normalsize

\subsection{Pairwise Velocity Moments}

Figure 13 shows the first ($v_{12}$) and the second moment ($\sigma_{12}$) of the one-dimensional radial 
pairwise velocity distribution of galaxies and dark matter particles as a function of pair separation
for our best-fit models.
Bold lines are the moments for the dark matter particles. Solid and dotted lines are the
moments for galaxies brighter than $M_B=-17.5 +
5\log h$ and $M_B=-18.5 + 5\log h$, respectively. Brighter cutoffs yield
similar results. Galaxies do not show any significant velocity bias on
any non-linear scale regardless of their luminosity. This result is
not trivial because galaxies are not a random subsample of dark matter particles.
This result is consistent with the weak density bias and the lack of
luminosity segregation described in Sect. 6.6.

The pairwise velocity dispersion profile $\sigma_{12}(r)$ is sensitive to the presence of
clusters within the sample and therefore to the total volume of the sample. Our simulation box
has a volume of the order of $10^6 h^{-3}$ Mpc$^3$ which is
comparable to the volume  $\sim 7\times 10^5 h^{-3}$ Mpc$^3$
within $cz=12000$ km s$^{-1}$ of the northern slice of the Center for Astrophysics Redshift
Surveys (hereafter CfA2N, de Lapparent et al 1991; Geller \& Huchra 1989; Huchra et al 1990;
Huchra et al 1995). Filled  squares in figure 13
show $\sigma_{12}(r)$ for the CfA2N slice. Both $\tau$CDM and
$\Lambda$CDM models are in remarkably good agreement with the CfA2N slice.

Lacking the full three-dimensional information, we compute the CfA2N values
by fitting the redshift space correlation function $\xi (r_p,\pi)$ with the
convolution model (Fisher 1995): we model $\xi (r_p,\pi)$ 
as the convolution of the real space correlation function with an 
exponential pairwise velocity distribution (Davis \& Peebles 1983; 
details of this procedure
are given in Marzke et al 1995). In order to apply this procedure,
we need to assume a model for the mean streaming velocity $v_{12}(r)$. We adopt the similarity
solution suggested by Davis \& Peebles (1983) $v_{12}(r)=r/[1+(r/r_0)^2]$ where 
$r_0=5.83h^{-1}$ Mpc is the CfA2N galaxy-galaxy correlation length. Assuming  $v_{12}(r)=0$
usually yields lower values of $\sigma_{12}$ (see e.g. Marzke et al 1995; Somerville, Primack 
\& Nolthenius 1997).
Note that our simulations show $v_{12}\sim 200-300$ km s$^{-1}$
at relative separations $r\sim 1h^{-1}$ Mpc and do not support the
assumption of a null $v_{12}(r)$.

In future papers, we use mock redshift surveys to show that this procedure 
yields pairwise velocity dispersion profiles $\sigma_{12}(r)$ in 
reasonably good agreement with the real profile.
It is therefore encouraging that the $\sigma_{12}(r)$ profiles of our best-fit models match
the CfA2N result. 

\begin{figure*}
\vbox to11.cm{\rule{0pt}{11.cm}}
\includegraphics{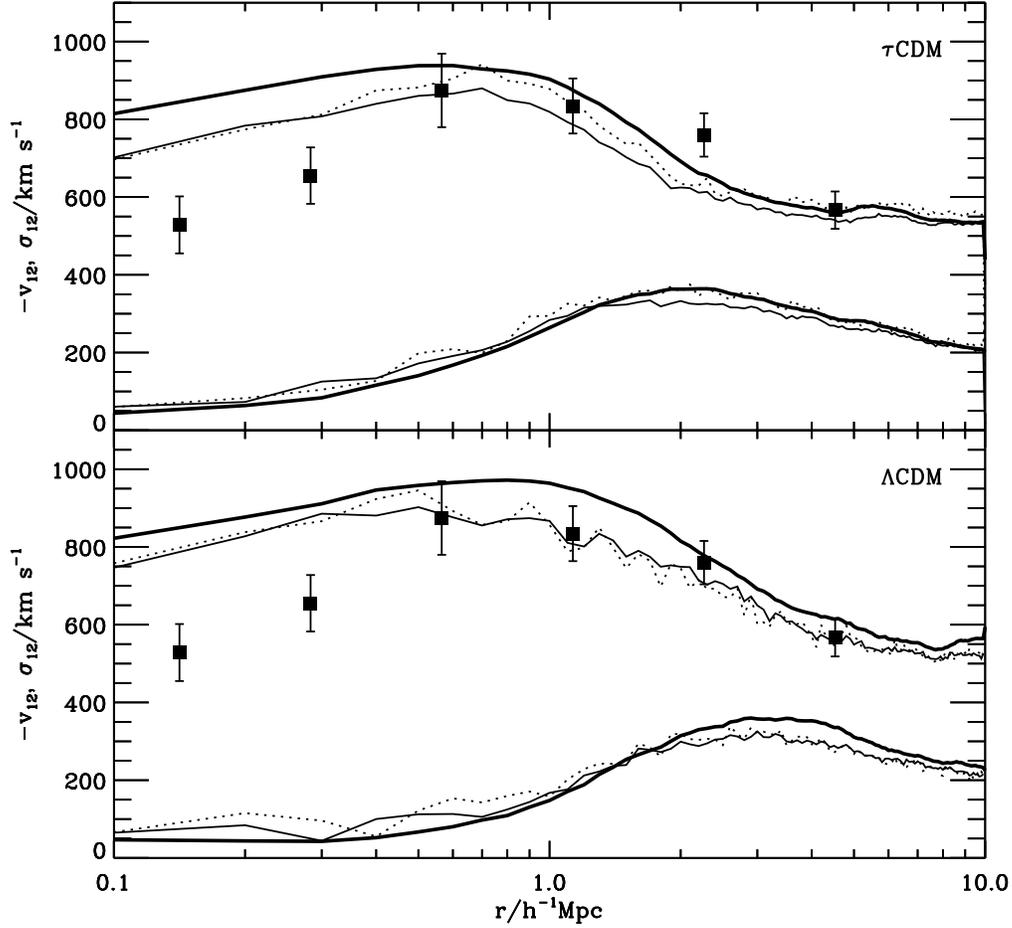}
\caption{
Mean streaming velocity $v_{12}$ (lower curves) and pairwise velocity dispersion
$\sigma_{12}$ (upper curves) at different relative separations for our best-fit models.
Bold, solid, and dotted lines are the moments for the dark matter particles,
galaxies brighter than $M_B=-17.5 +
5\log h$ and $M_B=-18.5 + 5\log h$, respectively. Solid squares are
the pairwise velocity dispersions for the CfA2N redshift survey. Error bars are
computed with a bootstrap resampling method.}
\label{fig:13}
\end{figure*}

\subsection {Galaxy Bias}
An important issue  we wish to address is that of galaxy ``bias'', i.e. whether the clustering amplitude
of the galaxies in our models differs from that of the underlying dark matter. In figure 14, we compare
$\xi(r)$ for galaxies of different luminosities, morphological types and colours with $\xi(r)$
for the dark matter. For $\tau$CDM, we show the model with ejection feedback
and dust extinction and for $\Lambda$CDM, we show the model with retention feedback and dust extinction.
These are the ``best fit'' models in both cases.      .

On large scales,  galaxies of all luminosities are unbiased in both models.
On small scales, the $\tau$CDM galaxies   
are slightly more clustered than the dark matter, whereas the $\Lambda$CDM galaxies are a factor $\sim 2-3$
less clustered. Neither model displays any  luminosity segregation; galaxies
of all magnitudes cluster in the same way. As discussed by Kauffmann, Nusser \& Steinmetz (1997), 
positive bias is obtained for galaxies in halos more massive than
$M_*(z=0)$, the mass of the typical collapsed object at the present day. For cluster-normalized
CDM models, $M_*$ is  large ($\sim 10^{14} M_{\odot}$). Our bright 
normalization also causes luminous galaxies to be placed in relatively low mass halos. This is why
we do not see the luminosity-dependent bias discussed by  Kauffmann, Nusser \& Steinmetz (1997).

On the other hand, red galaxies and to a lesser extent early-type galaxies are more clustered than the    
underlying population, particularly on small scales. These galaxies occur
preferentially in the cores of groups and clusters (see figures 4 and 5). 
Conversely, star-forming galaxies
are  less clustered on small scales because they are predominantly central halo galaxies and are
thus spatially exclusive. A similar result can be seen in the earlier work of Cen \& Ostriker (1992).

The relative bias between different types of galaxies in the Southern Sky Redshift Survey has recently
been  analyzed in a paper by Willmer, Da Costa \& Pellegrini (1998). These authors find that galaxies
with magnitudes below $\sim L_*$ do not exhibit any luminosity-dependent bias, but at magnitudes
brighter than $L_*$, galaxies are significantly more strongly clustered ($b(L>L_*)/b(L_*) = 1.5$).
As discussed above, our models do not exhibit this luminosity-dependent bias, perhaps indicating that our adopted
normalization is too bright.
Willmer, Da Costa \& Pellegrini also find that both early-type galaxies and red galaxies cluster more strongly than
the galaxy population as a whole. The bias is stronger on small scales ($r < 4$ $h^{-1}$ Mpc) than on larger
scales. In addition, galaxies with {\em colours} characteristic of old stellar populations
exhibit more bias than galaxies with early-type morphologies. These results are in good
qualitative agreement with the results shown in figure 14 and the 
earlier results of Kauffmann, Nusser \& Steinmetz (1997).
\begin{figure}
\centerline{
\epsfxsize=11cm \epsfbox{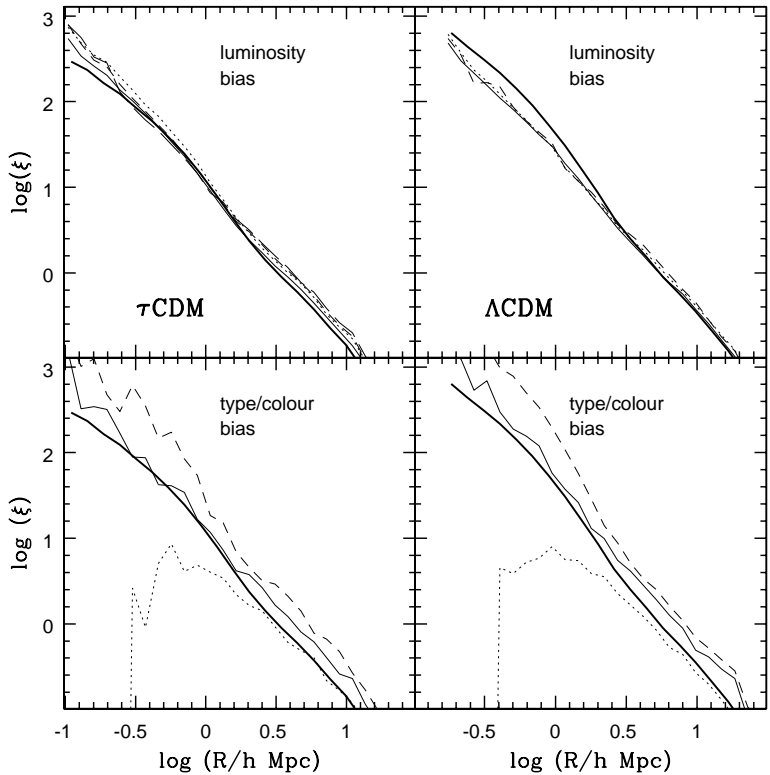}
}
\caption{\label{fig14}
\small
Galaxy bias as a function of luminosity, type, colour and star formation rate.
The left panels are for the $\tau$CDM simulation, the right panels are for the $\Lambda$CDM
simulation.
{\em Upper panels:} The thick solid line is the dark matter correlation function. Thin solid
lines show $\xi(r)$ for different B-band luminosities: $B<-19$ (solid), $B< -20$ (dotted),
$B < -20.5$ (short-dashed), $B < -21$ (long-dashed).
{\em Lower panels:} Thick solid line is the dark matter correlation function. Thin solid line
 corresponds to early-type galaxies, dashed line to red ($B-V >0.8$) galaxies, and dotted line to star forming
($SFR > 2 M_{\odot} \quad \rm{yr}^{-1}$) galaxies.}
\end {figure}
\normalsize

\subsection {Mass-to-light Ratios of Clusters}
Clusters of galaxies are the largest collapsed objects in the universe and have often been used as
a means of estimating the cosmological density parameter $\Omega$. The standard procedure is as follows.
The mass of a cluster within some radius is measured using techniques such as galaxy
kinematics, X-ray profiles or gravitational lensing. This is compared to the luminosity within the same radius
to calculate a mass-to-light ratio. The mean mass density of the universe is then
estimated by multiplying its mean luminosity density by this number.
The validity of this technique depends on two questionable assumptions:
\begin {enumerate}
\item Luminosity is ``conserved'' when field galaxies fall into a cluster. In practice, we know
  that cluster galaxies are predominantly ellipticals with no ongoing star formation, whereas
  field galaxies are predominantly spirals in which stars are forming at several solar
  masses per year. The stellar mass-to-light ratios of field and cluster galaxies  thus differ strongly 
  at optical and UV wavelengths.
\item The efficiency of galaxy formation is the same in all environments.
\end {enumerate}
Most studies of cluster mass-to-light ratios give values corresponding to  $\Omega \sim 0.2$ (see for example
Carlberg et al 1996). If a high-density cosmology is to be consistent with these observations, we  
require the mass-to-light ratios of clusters to be
significantly {\em lower} than the mass-to-light ratio of the Universe as a whole.    
This is often also
referred to as ``bias'' in the literature, though this kind of bias is quite different to
what we measure when we compare the correlation functions of galaxies and dark matter.

Figure 15 shows  the mass-to-light ratios of our simulated halos divided
by the mass-to-light ratio of the simulation as a whole. The results are shown as a function of
the virial mass of the  halo and for both B and I-bands. 
It is interesting that in the $\Omega=1$ $\tau$CDM model, the mass-to-light ratios of clusters
{\em are} significantly lower than the global value. The factor ($\sim 2$) that we obtain is somewhat too small,
however,
to bring the model into good  agreement with the observations. Rich clusters in the $\Lambda$CDM
model have mass-to-light ratios only 10-20 \% smaller than the global value. It is also interesting that groups
of galaxies are predicted to have lower mass-to-light ratios than clusters in both models. 
This agrees with the observed trends.

\begin{figure}
\centerline{
\epsfxsize=10cm \epsfbox{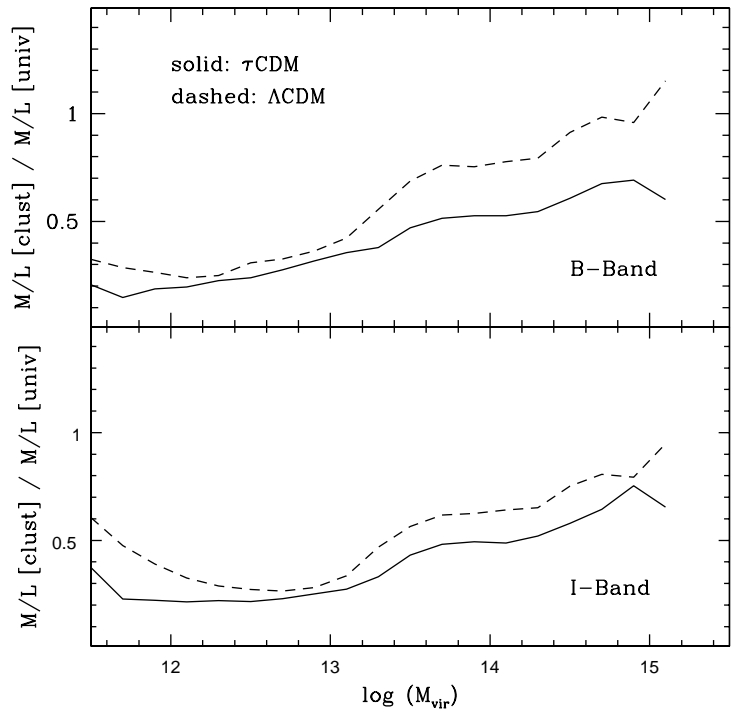}
}
\caption{\label{fig15}
The mass-to-light ratios of halos divided
by the mass-to-light ratio of the Universe as a whole. The results are shown as a function of
the virial mass of the  halo.} 
\small
\end {figure}
\normalsize

\subsection {Colour Distributions}
We now compare the $B-V$ colour distributions of our model galaxies with
colours of bright nearby galaxies in the RC3 catalogue (De Vaucouleurs et al 1991). We have selected  1580
galaxies with $M_B < -19.0-5\log h $, which have both morphological
classifications and magnitudes in the Johnson $B$ and $V$-bands. Of these galaxies, 653 are classified as
early-type. Note that the fraction of early-type galaxies is rather high, which may reflect the
fact that the sample is dominated by galaxies in the Virgo cluster. In our simulation $\sim 25\%$ of galaxies
to this magnitude limit have $M(B)_{bulge}-M(B)_{total} < 1$ mag and so are classified as early types.

In figure 16, we compare the colour distributions of galaxies in the
$\tau$CDM model with that of the RC3 sample. The left panel shows the model with ejection feedback,
but without any dust extinction. The right panel  shows the same model with dust extinction.
Without dust extinction, the model  colour distribution peaks at $B-V \sim 0.6$ in contrast to the
observed distribution, which is relatively flat from $B-V \sim 0.4$ to $B-V \sim 1$. In particular, early-type
galaxies appear to be significantly too blue in this model. With the inclusion of dust, the color distribution
broadens significantly. Many of the field ellipticals 
on our models have a lower-luminosity, star-forming  disk component. The inclusion of dust in the models
reddens these objects and brings the colour distribution of early-type galaxies into better agreement with
the observations.

\begin{figure}
\centerline{
\epsfxsize=12cm \epsfbox{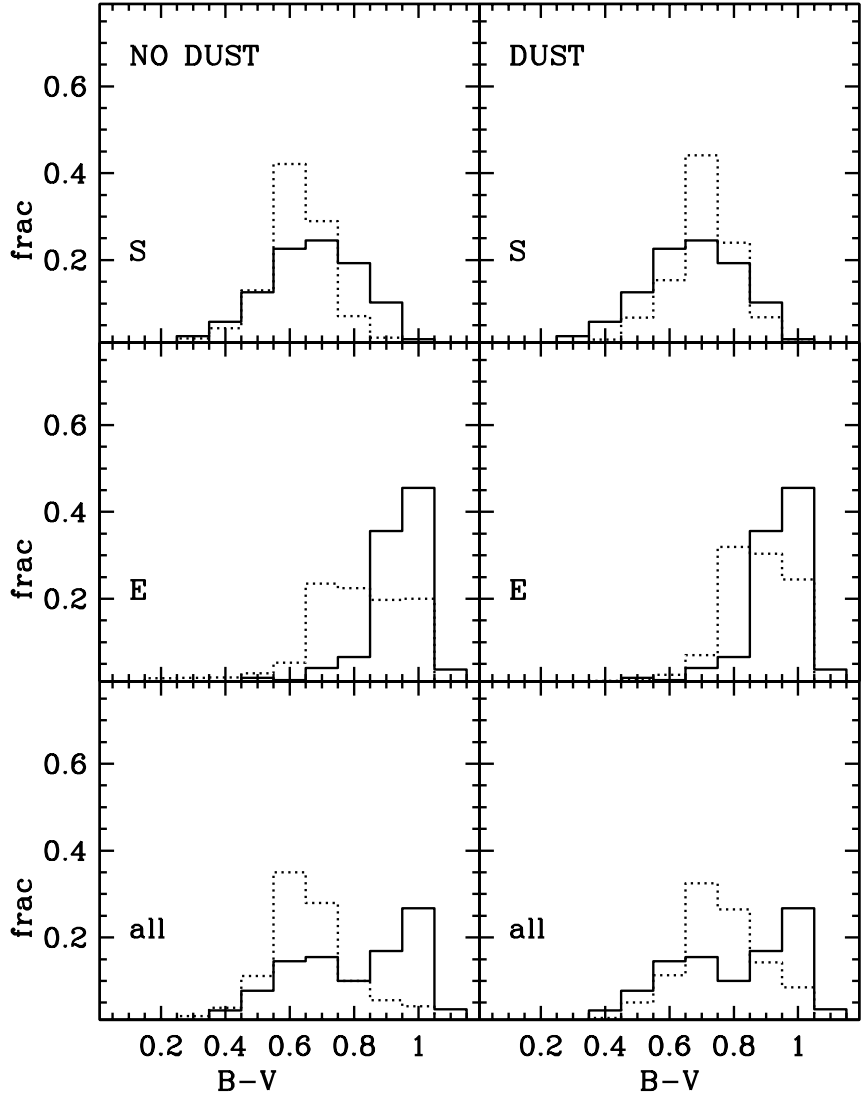}
}
\caption{\label{fig16}
The $B-V$ colour distribution of galaxies brighter than $B=-20.5$ in the $\tau$CDM simulation.
The left panel shows a model without dust and the right panel  a model with dust.
Dotted lines show the model distributions. Solid lines show the distribution of $B< -20.5$ galaxies
with both colours and morphologies in the RC3 catalogue.}
\small
\end {figure}
\normalsize

The colour distributions of galaxies in the $\Lambda$CDM models are shown in figure 17. Because galaxies
form earlier in this cosmology, the colours tend to be redder on average, but the effect is small.
Again, the
effect of dust extinction is to broaden the colour distributions and to bring them into better agreement
with the observations. 

\begin{figure}
\centerline{
\epsfxsize=12cm \epsfbox{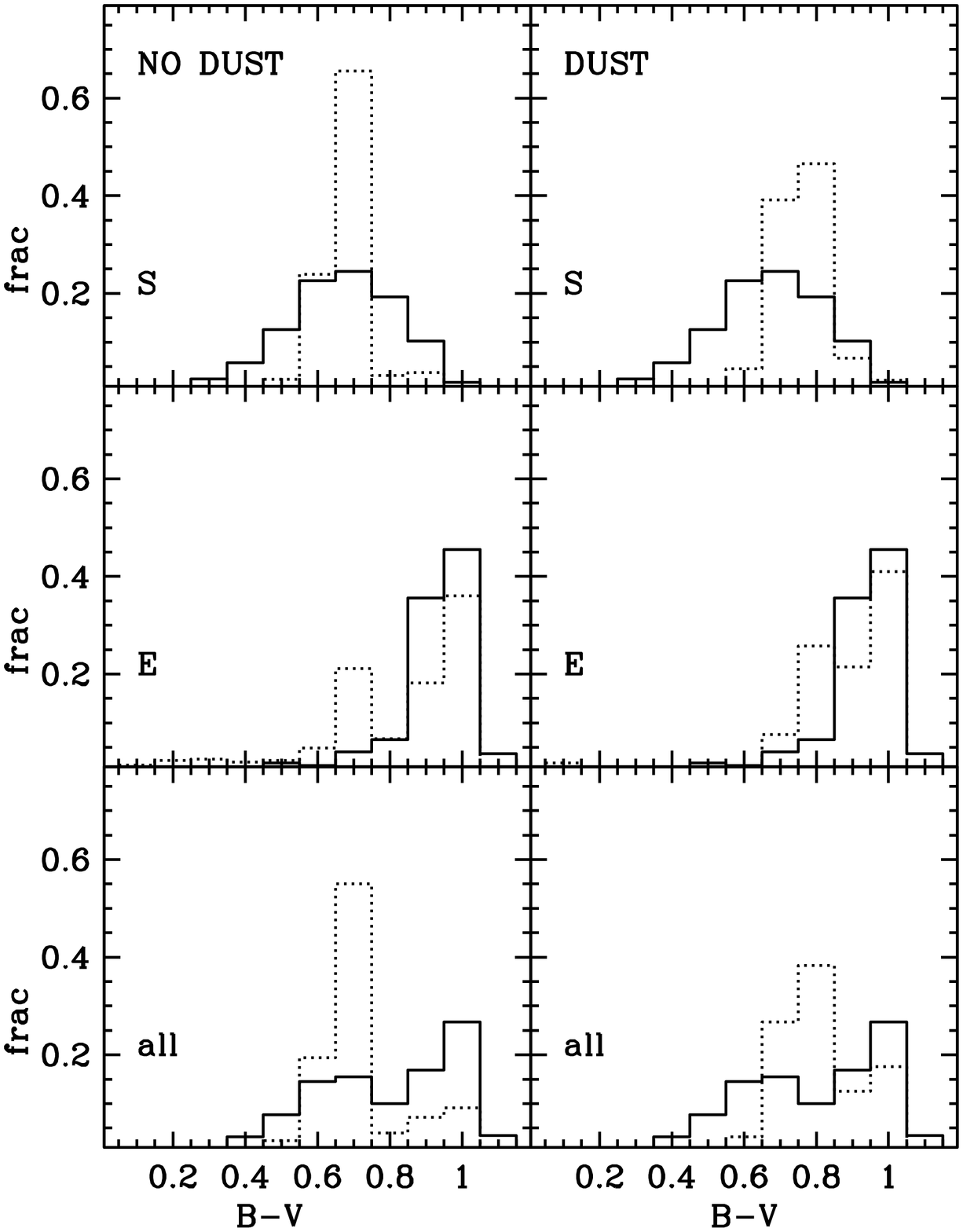}
}
\caption{\label{fig17}
The $B-V$ colour distribution of galaxies brighter than $B=-19.7$ in the $\Lambda$CDM simulation.
The left panel shows a model without dust and the right panel a model with dust.}
\small
\end {figure}
\normalsize

\subsection {Star Formation Rate Functions}
In addition to the luminosity function and the colour distributions of galaxies, observational estimates of the
number density of galaxies as a function of star formation rate provide an important constraint on
galaxy formation theories. Gallego et al (1995) have analyzed a complete
sample of emission-line galaxies and have computed an H$\alpha$ luminosity function. Assuming a Scalo (1986)
IMF and using Case B recombination theory to predict the luminosity of the H$\alpha$ emission line
for a given rate of star formation, they derive an integrated star formation rate density of the
local Universe of $0.013 \pm 0.001 M_{\odot} \rm{yr}^{-1} $ Mpc $^{-3}$ (for $H_0= 50$ km s$^{-1}$ Mpc$^{-1}$).

In figure 18, we compare the H$\alpha$ luminosity functions of our models with the results of
Gallego et al. We have used the same transformation as these authors to convert from star formation rate to
H$\alpha$ luminosity,
\begin {equation} L(H\alpha) = 9.40 \times 10^{40} \frac {SFR} {M_{\odot} \rm{yr}^{-1}} \rm{ergs} \quad \rm{ s}^{-1}. \end
{equation}
The results of the $\tau$CDM model agree   better with the observations
than those of the $\Lambda$CDM model. For $\tau$CDM, we obtain an integrated SFR density of 0.017
$M_{\odot}$ yr$^{-1}$ Mpc$^{-3}$. For $\Lambda$CDM the
present-day SFR density is a factor $\sim 4$ too low. From figure 18, we see that there
there is a lack of galaxies forming stars at rates between 1 and 10 $M_{\odot}$ yr$^{-1}$.
These are the same galaxies that are missing at the ``knee'' of the B-band luminsity function
in figure 9.

\begin{figure}
\centerline{
\epsfxsize=9cm \epsfbox{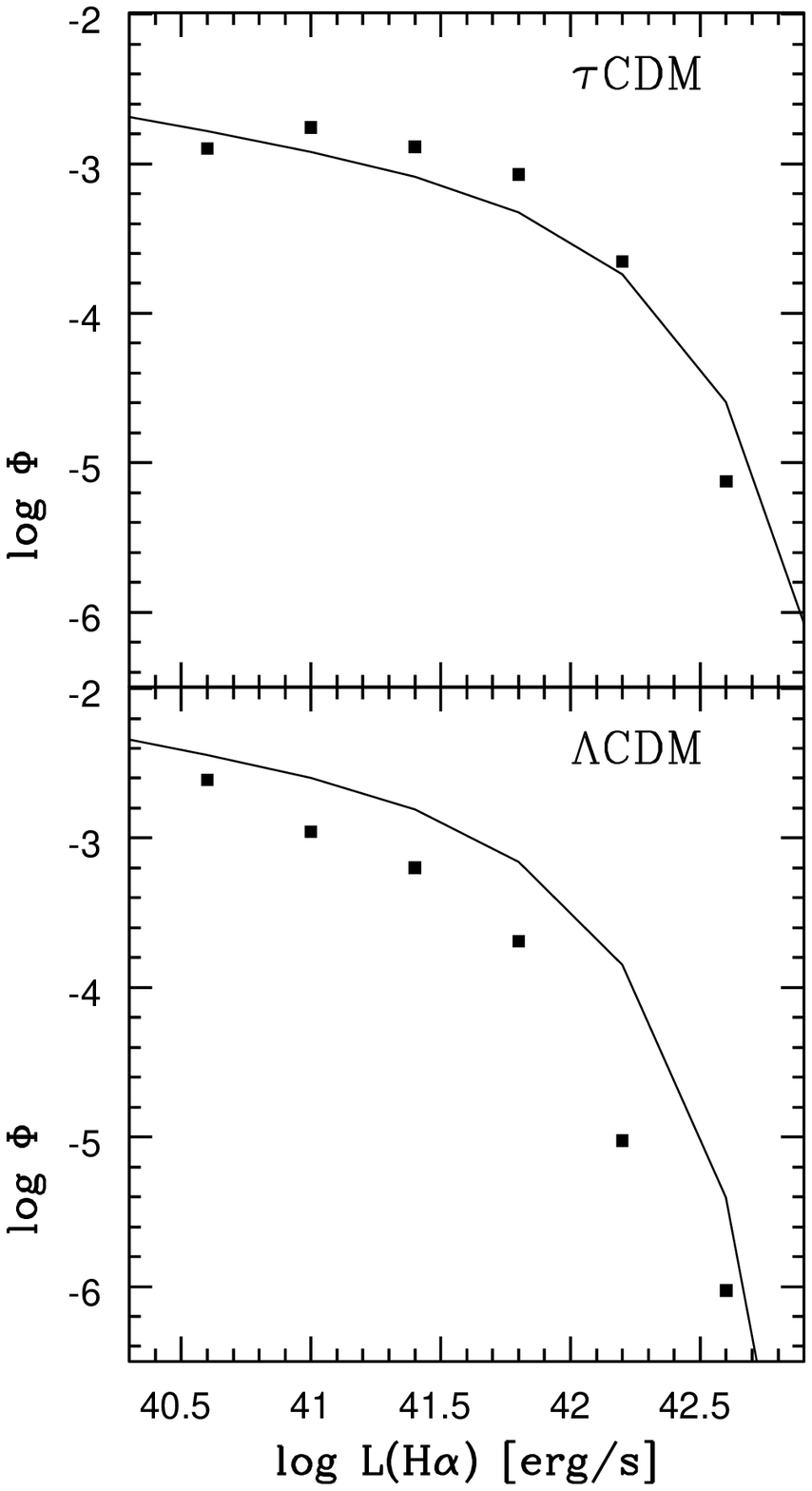}
}
\caption{\label{fig18}
The H$\alpha$ luminosity function of galaxies in the simulation compared to the Schechter fit derived
by Gallego et al (1995).} 
\small
\end {figure}
\normalsize

\section {Discussion and Conclusions}

None of the models analyzed in this paper provides a completely
satisfactory fit to all the observed properties
of galaxies at $z=0$. If supernova feedback is weak,  
the $\Omega=1$ $\tau$CDM model produces too high a total luminosity density.
Most of the excess luminosity is in the form of star-forming ``field'' galaxies. 
As a result, the galaxy correlation function turns over               
on scales below a few Mpc. The observed  two-point correlation
function, on the other hand,  is well represented by a single  
power law of slope $\ -1.7$ down to scales approaching 10 kpc (Baugh 1996).
If one assumes instead  that supernova feedback is able to eject gas
out of a halo, so that it becomes unavailable for further cooling and star formation
until much later times, the number of star-forming
galaxies in low-mass field halos is significantly  lowered and this improves 
the fit to the observed luminosity and correlation functions.
Even so, the model still produces an excess of
very bright galaxies in groups and clusters.
The $\Omega=0.3$ $\Lambda$CDM model has the opposite problem: it produces too few $L_*$ galaxies,      
even if feedback is weak and no gas leaves the halo.
The two point correlation function is also too steep on scales
below a few Mpc. 

One obvious inference we can draw from these
results is that a CDM model with density parameter somewhere between $0.3$ and $1$ may well 
do  better in matching the observations. Indeed, in recent work,  Somerville \& Primack (1998)
find that a model with $\Omega=0.5$ provides the best fit.

Care must be exercised, however, before drawing strong conclusions about cosmological parameters.
Many of our difficulties in fitting the observed shape of the galaxy luminosity function come about
because we have chosen a very bright normalization for the models. This is because we {\em force}
the models to fit the zero-point of the Giovanelli et al I-band Tully-Fisher relation at a circular velocity of 
$V_c = 220$ km s$^{-1}$. We have also assumed that the circular velocities of central disk galaxies are
the same as those of their surrounding halos. In practice, the relation between the circular velocities
of disk and  halo is not straightforward and requires a detailed model for the internal structure of the
halo and of the galaxy within it.                                                             
More detailed calculations of disk formation in CDM halos                                    
predict that the circular velocity of the disk is generally {\em higher}
than that of the surrounding halo (Mo, Mao \& White 1998). This results in a higher total luminosity density 
for the same Tully-Fisher normalization, and would improve the fit of the $\Lambda$CDM model while
making the $\tau$CDM model even more discrepant.    

One might also question whether normalizing the models to match the Tully-Fisher relation is sensible,         
since the physical origin of this relation is still uncertain and depends on  structural properties
of the galaxies which we are not attempting to model.
We have argued that the distribution of  K-band luminosities should provide  a robust test
of galaxy formation theory, since the K-band luminosity of a galaxy depends primarily on its stellar mass and   
not on its instantaneous star formation rate or its  dust content. With the advent of wide-field K-band
galaxy surveys such as the Two Micron All Sky Survey (2MASS, Kleinmann et al 1994),
normalizing the models to match the observed K-band 
luminosity function may be the best choice for the future.

Another effect we have investigated is that of dust extinction in star-forming galaxies.
If dust is ignored, the $\Omega=1$ $\tau$CDM model produces too many
blue galaxies compared to the observations. Galaxy formation occurs rather
late in this model, so the stellar populations of galaxies are younger than in the
$\Lambda$CDM model. However,  the inclusion of a simple, empirically-motivated
dust extinction scheme makes  this problem go away, and  
the colour distributions of galaxies in the $\tau$CDM and $\Lambda$CDM models 
are then virtually indistinguishable.
Since dust extinction only affects star-forming field galaxies, models with dust produce somewhat
steeper and higher amplitude correlation functions.
Finally, dust lowers the predicted number of galaxies at the bright end of the B-band luminosity
function. Our models tend to produce a  power-law  ``tail''  of high-luminosity galaxies, rather
than a sharp exponential cutoff. Although dust can reduce this tail in the B-band, the problem
persists in the K-band.

Finally, although we have tried to illustrate the effect of different schemes for star formation, feedback
 and dust, we cannot be sure that we have taken into account all the
key physical processes that may seriously affect our results. For example, we have proposed that
very efficient feedback can stop galaxies forming in low-mass halos, but alternative schemes may be
possible. 
For example, Jimenez et al (1997) have proposed that the gaseous disks that  form in low $V_c$ halos are not subject
to the instabilities that cause star formation and are thus not seen in the optical or UV,  
although their emission at  21 cm wavelengths should be detectable. The only way to make progress
is through observational studies of how
 star formation rates,  cold gas and dust contents,
and the shells and bubbles seen around star-forming regions, vary
with galactic mass, metallicity, morphology and environment. This is a massive and long-term undertaking,
but it seems necessary if we are to build a picture of galaxy formation that is more than just
schematic.

We conclude that we cannot reliably constrain the values of cosmological parameters
using the properties of the galaxy distribution at $z=0$ because  our results are strongly influenced
by our adopted normalization and the way in which we choose to parametrize star formation
and supernova feedback.
The main effects of changing the cosmological parameters in our set of cluster-normalized  CDM models are
\begin {enumerate}
\item to shift the epoch when halos of a given mass form  
\item to change the number density of galactic-mass halos at the present day.
\end {enumerate}
The adopted normalization and the recipes for feedback and star formation affect     
the efficiency with which stars form as a function of redshift and galaxy mass.
It is thus not surprising that there is considerable
freedom to ``tune'' our results to fit the observations. 
Nevertheless it is interesting that very different 
star formation and feedback schemes are required to come close to reproducing the
observations in the two cosmologies we have explored.
For $\Omega=1$, we require that feedback be very efficient at removing gas both from 
galaxies and from their surrounding halos. Many lower-mass halos are ``dark'' in this model, containing
a faint, possibly low-surface brightness  object. In a low-density Universe, feedback must be 
inefficient and every halo must contain a star-forming galaxy if we are to match the observations.
This gives us confidence that if the cosmological parameters of the Universe can be determined
using techniques such as the analysis of 
small-scale fluctuations in the microwave background (e.g. Bond, Efstathiou \&
Tegmark 1997)
or the light-curves of high-redshift supernovae 
(e.g. Perlmutter et al 1997), the methodology introduced in this paper
will become valuable for understanding and constraining the detailed physical processes operating
within galaxies.

It is also encouraging that our best-fit models come reasonably
close to fitting many different observed measures of the galaxy distribution  {\em at the same time}.          
The models that give the best fits to the B-band luminosity function, also give the best
fits to the K-band luminosity function, the correlation function, the color distribution
and the star formation rate function. In Paper II we demonstrate that these same models
provide a reasonably good match to the observed properties of galaxy groups.
This gives us confidence that the fundamental  theoretical
framework, which determines how galaxy properties  scale with halo mass,        
may indeed be correct. 

\vspace{0.8cm}

\large
{\bf Acknowledgments}\\
\normalsize
The simulations in this paper were carried out at the Computer Center of the Max-Planck Society 
in Garching and at the EPPC in Edinburgh. Codes were kindly made available by the Virgo Consortium.
We especially thank Adrian  Jenkins and Frazer Pearce for help in carrying them out.
We are also grateful to Margaret Geller, John Huchra and Ron Marzke for enabling us
to compare the CfA2N data directly with the simulations.
A.D. is a Marie Curie Fellow and holds grant ERBFMBICT-960695 of the Training and Mobility of
Researchers program financed by the EC.

\pagebreak

\begin{table}
\caption{Cosmological parameters of the GIF models. $\Omega_0$ and
  $\Omega_\Lambda$ are the density parameters for matter and for the
  cosmological constant, $h$ is the Hubble parameter, $\sigma_8$ is
  the rms of the density field fluctuations in spheres of radius $8\,h^{-1}\;$Mpc,
  and $\Gamma$ is the shape parameter of the power spectrum. Also
  given is the size of the cosmological simulation box.}
\label{tab:2.1}
\medskip
\centering
\begin{tabular}{l*{7}{c}}
\hline
Model & $\Omega_0$ & $\Omega_\Lambda$ & $h$ & $\sigma_8$ &
$\Gamma$ & Box Size  \\
& & & & & & [${\rm Mpc}/h$]  \\
\hline
SCDM         & 1.0 & 0.0 & 0.5 & 0.60 & 0.50 &  85  \\
$\tau$CDM    & 1.0 & 0.0 & 0.5 & 0.60 & 0.21 &  85  \\
$\Lambda$CDM & 0.3 & 0.7 & 0.7 & 0.90 & 0.21 & 141  \\
OCDM         & 0.3 & 0.0 & 0.7 & 0.85 & 0.21 & 141  \\
\hline
\end{tabular}
\end{table}

\pagebreak
\Large
\begin {center} {\bf References} \\
\end {center}
\normalsize
\parindent -7mm
\parskip 3mm

Allen, S.W. \& Fabian, A.C., 1997, MNRAS, 286, 583

Barnes, J.E. \& Hernquist, L., 1996, ApJ, 471, 155

Bartelmann, M., Huss, A., Colberg, J.M., Jenkins, A. \& Pierce, F.R., 1998, A\&A, 330, 1 

Baugh, C.M., Cole, S. \& Frenk, C.S., 1996a, MNRAS, 282, 27

Baugh, C.M., Cole, S. \& Frenk, C.S., 1996b, MNRAS, 283, 1361

Baugh, C.M., 1996, MNRAS, 280, 267

Baugh, C.M., Cole, S., Frenk, C.S. \& Lacey, C., 1998, ApJ, 498, 504

Binney, J. \& Tremaine, S., 1987, Galactic Dynamics, Princeton University Press

Bond, J.R., Efstathiou, G. \& Tegmark, M., 1997, MNRAS, 291, 33

Bond, J.R., Cole, S., Efstathiou, G. \& Kaiser, N. 1991, ApJ, 379, 440

Bower, R.J., 1991, MNRAS, 248, 332

Bower, R.G., Coles, P., Frenk, C.S. \& White, S.D.M., 1993. ApJ, 405, 403

Cardelli, J.A., Clayton, G.C. \& Mathis, J.S., 1989, ApJ, 345, 245

Carlberg, R.G. Couchman, H.M.P. \& Thomas, P., 1990, ApJ, 352, L29

Carlberg, R.G., Yee, H.K.C., Ellingson, E., Abraham, R., Gravel, P., Morris, S.
\& Pritchet, C.J., 1996, ApJ, 462, 32

Cen, R. \& Ostriker, J.P., 1992, ApJ, 393, 22

Cen, R. \& Ostriker, J.P., 1993, ApJ, 417, 415

Cole, S., 1991, ApJ, 367, 45

Cole, S., Arag\'on-Salamanca, A., Frenk, C.S., Navarro, J.F.
\& Zepf, S.E., 1994, MNRAS, 271, 781

Couchman, H.M.P., Thomas, P.A. \& Pearce, F.R., 1995, ApJ, 452, 797

Davis, M. \& Peebles, P.J.E., 1983, ApJ, 267, 465

De Jong, S., 1996, A\&AS, 118, 557

De Lapparent, V., Geller, M.J. \& Huchra, J.P., 1991, ApJ, 369, 273

De Vaucouleurs, G., De Vaucouleurs, A., Corwin, H.G., Buta, R.J., Paturel, G. \& Fouque, P., 1991,
Third Reference Catalogue of Bright Galaxies, Springer-Verlag

Eke, V.R., Cole, S. \& Frenk, C.S., 1996, MNRAS, 282, 263

Evrard, A.E., Summers, F.J. \& Davis, M., 1994, ApJ, 422, 11

Fabian, A.C., Nulsen, P.E.J. \& Canizares, C.R., 1991, A\&ARv, 2, 191

Ferland, G.J., Fabian, A.C. \& Johnstone, R.M., 1994, MNRAS, 266, 399

Fisher, K.B., 1995, ApJ, 448, 494

Frenk, C.S., Evrard, E.E., White, S.D.M. \& Summers, F.J., 1996, ApJ, 472, 460

Gallego, J., Zamorano, J., Aragon-Salamanca,A. \& Rego, M., 1995, ApJ, 455, L1

Gardner, J.P., Sharples, R.M., Frenk, C.S. \& Carrasco, B.E., 1997, ApJ, 480, 99

Geller, M.J. \& Huchra, J.P., 1989, Science 246, 897

Ghigna, S., Moore, B., Governato, F., Lake, G., Quinn, T. \& Stadel, J., 1998, MNRAS, in press

Giovanelli, R.M., Haynes, M.P., Da Costa, L.N., Freudling, W., Salzer, J.J. \&
Wegner, G., 1997, ApJ, 477, L1

Governato, F., Baugh, C.M. , Frenk, C.S. , Cole, S., Lacey, C.G., Quinn, T. \& Stadel, J., 1998, Nature,
392, 359

Heyl, J.S., Cole, S., Frenk, C.S. \& Navarro, J.F., 1995, MNRAS, 274, 755

Huchra, J.P., de Lapparent, V., Geller, M.J. \& Corwin, H.G., 1990, ApJS, 72, 433

Huchra, J.P., Geller, M.J. \& Corwin, H.G., 1995, ApJS, 70, 687

Jimenez, R., Heavens, A.F., Hawkins, M.R.S. \& Padoan, P., 1997, MNRAS, 292, L5

Katz, N., Hernquist, L. \& Weinberg, D.H., 1992, ApJ, 399, L109

Kauffmann, G. \& White, S.D.M., 1993, MNRAS, 261, 921

Kauffmann, G., White, S.D.M. \& Guiderdoni, B. 1993, MNRAS, 264, 201 (KWG)

Kauffmann, G., Guiderdoni, B. \& White, S.D.M., 1994, MNRAS, 267, 981

Kauffmann, G., 1995a, MNRAS, 274, 153

Kauffmann, G., 1995b, MNRAS, 274, 161

Kauffmann, G., 1996a, MNRAS, 281, 487

Kauffmann, G., 1996b, MNRAS, 281, 475

Kauffmann, G., Nusser, A. \& Steinmetz, M., 1997, MNRAS, 286, 795

Kauffmann, G. \& Charlot, S., 1998, MNRAS, 294, 705

Kennicutt, R.C., 1998, ApJ, 498, 541

Kleinman, S.G., Lysaght, M.G., Pughe, W.L., Schneider, S., Skrutskie, M.F.,
Weinberg, M.D., Price, S.D., Matthews, K., Soifer, B.T. \& Huchra, J.P., 1994, Ap\&SS, 217, 11

Klypin,A.A., Gottloeber, S., Kravtsov, A.V. \& Khokhlov, A.M., 1998, ApJ, submitted

Lacey, C. \& Silk, J., 1991, ApJ, 381, 14

Lacey, C. \& Cole, S., 1993, MNRAS, 262, 627

Lacey, C., Guiderdoni, B., Rocca-Volmerange, B. \& Silk, J., 1993, ApJ, 402, 15

Lin, H., Kirshner, R.P., Schechtman, S.A., Landy, S.D., Oemler, A., Tucker, D.L. 
\& Schechter, P.L., 1997, ApJ, 464, 60

Loveday, J., Peterson, B.A., Efstathiou, G. \& Maddox, S.J., 1992, ApJ, 390, 338 

Marzke, R.O., Geller, M.J., Da Costa, L.N. \& Huchra, J.P., 1995, AJ, 110, 477

Mo, H.J. \& White, S.D.M., 1996, MNRAS, 282, 347

Mo, H.J., Mao, S. \& White, S.D.M., 1998, MNRAS, 295, 319  

Navarro, J.F. \& White, S.D.M., 1994, MNRAS, 267, 401

Navarro, J.F., Frenk, C.S. \& White, S.D.M., 1995, MNRAS, 275, 56

Navarro, J.F., Frenk, C.S. \& White, S.D.M., 1996, ApJ, 462, 563

Peacock, J.A. \& Dodds, S.J., 1994, MNRAS, 267, 1020

Pearce, F.R. \& Couchman, H.M.P., 1997, New Astronomy, 2, 411    

Perlmutter, S., Gabi, S., Goldhaber, G., Goobar, A., Groom, D.E., Hook, I.M.,
Kim, A.G., Kim, M.Y. et al., 1997, ApJ, 484, 565

Rodrigues, D.D.C. \& Thomas, P.A., 1996, MNRAS, 282, 631

Roukema, B.F., Peterson, B.A., Quinn, P.J. \& Rocca-Volmerange, B., 1997, MNRAS, 292, 835

Scalo, J.M., 1986, Fundamentals of Cosmic Physics, 11, 3

Simien, F. \& De Vaucouleurs, G., 1986, ApJ, 302, 564

Somerville, R.S., Davis, M. \& Primack, J.R., 1997, ApJ, 479, 616

Somerville, R.S., Primack, J.R. \& Nolthenius, R., 1997, ApJ, 479, 606

Somerville, R.S. \& Kolatt, T., 1998, MNRAS, in press

Somerville, R.S. \& Primack, J.R., 1998, MNRAS, submitted

Szokoly, G.P., Subbarao, M.U., Connolly, A.J. \& Mobasher, B., 1998, ApJ, 492, 452

Tormen, G., 1998, MNRAS, submitted

Tormen, G., Diaferio, A. \& Syer, D., 1998, MNRAS, in press

Van Kampen, E. \& Katgert, P., 1997, MNRAS, 289, 327

Viana, P.T.P. \& Liddle, A.R., 1996, MNRAS, 281, 323

Wang, B. \& Heckman, T.M., 1996, ApJ, 457, 645

White, S.D.M., Davis, M., Efstathiou, G. \& Frenk, C.S., 1987, Nature, 330, 451

White, S.D.M. \& Frenk, C.S., 1991, ApJ, 379, 52

White, S.D.M., Efstathiou, G. \& Frenk, C.S.,1993, MNRAS, 262, 1023

White, M., Gelmini, G. \& Silk, J., 1995, Phys.Rev.D, 51, 2669

Willmer, C.N.A., Da Costa. N. \& Pellegrini, P.S., 1998, AJ, 115, 869

Zucca, E., Zamorani, G., Vettolani, G., Cappi, AS., Merighi, R., Mignoli, M.,
Stirpe, G.M., Macgillivray, H. et al., 1997, A\&A, 326, 477

\end {document}